\documentclass[superscriptaddress, twocolumn, prb, bibnotes, aps, longbibliography, floatfix]{revtex4-2}
\usepackage{comment, siunitx}
\usepackage[utf8]{inputenc}
\usepackage{graphicx}
\usepackage{afterpage}
\usepackage{amsmath, amssymb, bbm, bm}
\usepackage{xcolor}
\usepackage{float,placeins}

\newcommand{\bmh}[1]{\bm{\hat #1}}

\begin{document}

\title{Topological supermodes in photonic crystal fiber}

\author{Nathan Roberts}
\affiliation{Department of Physics, University of Bath, Claverton Down, Bath BA2 7AY, UK}
\affiliation{Centre for Photonics and Photonic Materials, University of Bath, Bath BA2 7AY, UK}

\author{Guido Baardink }

\affiliation{Department of Physics, University of Bath, Claverton Down, Bath BA2 7AY, UK}

\author{Josh Nunn}
\affiliation{Department of Physics, University of Bath, Claverton Down, Bath BA2 7AY, UK}
\affiliation{Centre for Photonics and Photonic Materials, University of Bath, Bath BA2 7AY, UK}
\affiliation{ORCA Computing Ltd., 30 Eastbourne Terrace, London W2 6LA, UK}

\author{Peter J.~Mosley}
\email{p.mosley@bath.ac.uk}
\affiliation{Department of Physics, University of Bath, Claverton Down, Bath BA2 7AY, UK}
\affiliation{Centre for Photonics and Photonic Materials, University of Bath, Bath BA2 7AY, UK}

\author{Anton Souslov}
\email{a.souslov@bath.ac.uk}
\affiliation{Department of Physics, University of Bath, Claverton Down, Bath BA2 7AY, UK}
\begin{abstract}
   Topological states enable robust transport within disorder-rich media through integer invariants inextricably tied to the transmission of light, sound, or electrons. However, the challenge remains to exploit topological protection in a length-scalable platform such as optical fibre. We demonstrate, through both modelling and experiment, optical fibre that hosts topological supermodes across multiple light-guiding cores.
We directly measure the photonic winding-number invariant characterising the bulk and observe topological guidance of visible light over metre length scales. Furthermore, the mechanical flexibility of fibre allows us to reversibly reconfigure the topological state. As the fibre is bent, we find that the edge states first lose their localization and then become relocalised due to disorder. We envision fibre as a scalable platform to explore and exploit topological effects in photonic networks
\end{abstract}
\maketitle
\section*{Introduction}
Inspired by electronic quantum materials~\cite{Hasan2010}, metamaterial structures have recently been used to engineer topological bands for photonic~\cite{Raghu2008analogs,Wang2009observation,Rechtsman2013,Khanikaev2013,Peterson2018quantized,Shalaev2019,Ozawa2019} and acoustic~\cite{Nassar2020} systems.
These bands can be moulded to provide back-scattering-free propagation~\cite{Wang2009observation,Khanikaev2013}, disorder-resistant band gaps~\cite{Rechtsman2013}, and protected corner states~\cite{Peterson2018quantized}. 
Exploiting these phenomena in photonic chips has enabled topologically robust lasing modes~\cite{Harari2018topological,Bandres2018topological} and protected conduits for entangled quantum states~\cite{Blanco-Redondo2020,Kim2020,Duenas2021,Bergamasco2019,Ren:22}. 

Topological modes  are protected because of the topological invariant that characterises the system. This integer invariant can only change upon closing a band gap. By directly associating the invariant with physical characteristics such as light propagation,  topology can be used to endow a system with protection against fabrication-induced imperfections. The presence of topological modes relies on a few key properties, including the presence of a  band gap and the symmetries of the system. By selecting the necessary ingredients and structure, topological physics can be used to design mode profiles~\cite{Ren:22,Blanco-Redondo2020},  robust pumping~\cite{Zilberberg2018, Kraus2012}, and unidirectional transport~\cite{Chang2013, Wang2009observation}.

For microwaves, topology can be achieved through macroscale structures in combination with resonance effects~\cite{Wang2009observation,Khanikaev2013,Peterson2018quantized}. 
However, at optical frequencies, the challenge remains to create waveguides more than a few wavelengths long in which light is both confined by a micron-scale structure and topologically protected. 
For example, arrays of silicon waveguides are limited by scattering loss in the near infra-red~\cite{Blanco-Redondo2016}, whereas for visible light, state-of-the-art planar fabrication techniques, such as using lasers to inscribe waveguides into a glass chip, are challenging to extend beyond the centimetre scale~\cite{Wang2019}. 
Optical fibre that supports topological modes has previously been proposed~\cite{Lu2018topological,Lin2020,Pilozzi2020, Gong2021,Makwana2020hybrid}, but experimental demonstrations have not been seen due to the impossibility of fabricating previous designs with current technology.

Topological fibre could play a crucial role in next-generation quantum networks. Optical fibre is a key enabling technology for long-haul telecommunications networks due to its exceptionally low attenuation~\cite{Kato1973}.  In the drive to scale up quantum networks, this low attenuation also plays a key role due to the impossibility of noise-free amplification of quantum states~\cite{Kouznetsov1995,Josse2006}. In chip-scale experiments, light propagating in topologically protected modes over hundreds of microns to millimetres has been shown to inherit robustness from the topological medium~\cite{Blanco-Redondo2020,Kim2020,Duenas2021,Bergamasco2019,Bandres2018topological}. Therefore, if the guidance of topological modes could be demonstrated in fibre at length scales of metres to kilometres, this would offer topological robustness within distributed classical and quantum networks. In addition, topological fibre design has the potential to protect the generation of entangled states of light from the detrimental effects of fabrication-induced variations~\cite{Francis-Jones2016}.
 
Unlike telecommunications fibre, photonic crystal fibre (PCF) controls the propagation of light using a cross-section with wavelength-scale micro-structure~\cite{Knight2003,Russell2003}.
Tailoring the design of PCF allows unparalleled customisation of its dispersion and mode spectrum~\cite{Birks1997}. 
For example, many solid glass cores can be embedded in a periodic cladding of air holes to form a single flexible fibre with a typical diameter of 100--\SI{250}{\micro \metre}.
In this fibre, the individual modes of each core can overlap and collectively form so-called supermodes. 
This type of multi-core structure forms the platform for our topological PCF (TopoPCF) that translates the strategies for topological band engineering into fibre. Fabricating our TopoPCF allows us to observe propagation of edge states with visible light and to use bending for topological mode control. We show that bending the fibre leads to an effective
disorder inside our topological state, which we can switch on and off through mechanical reconfiguration. This unique control of topology via disorder exemplifies fibre-based phenomena that are inaccessible in planar waveguide architectures.

\begin{figure}[tbp]
\centering
\includegraphics[width=\columnwidth]{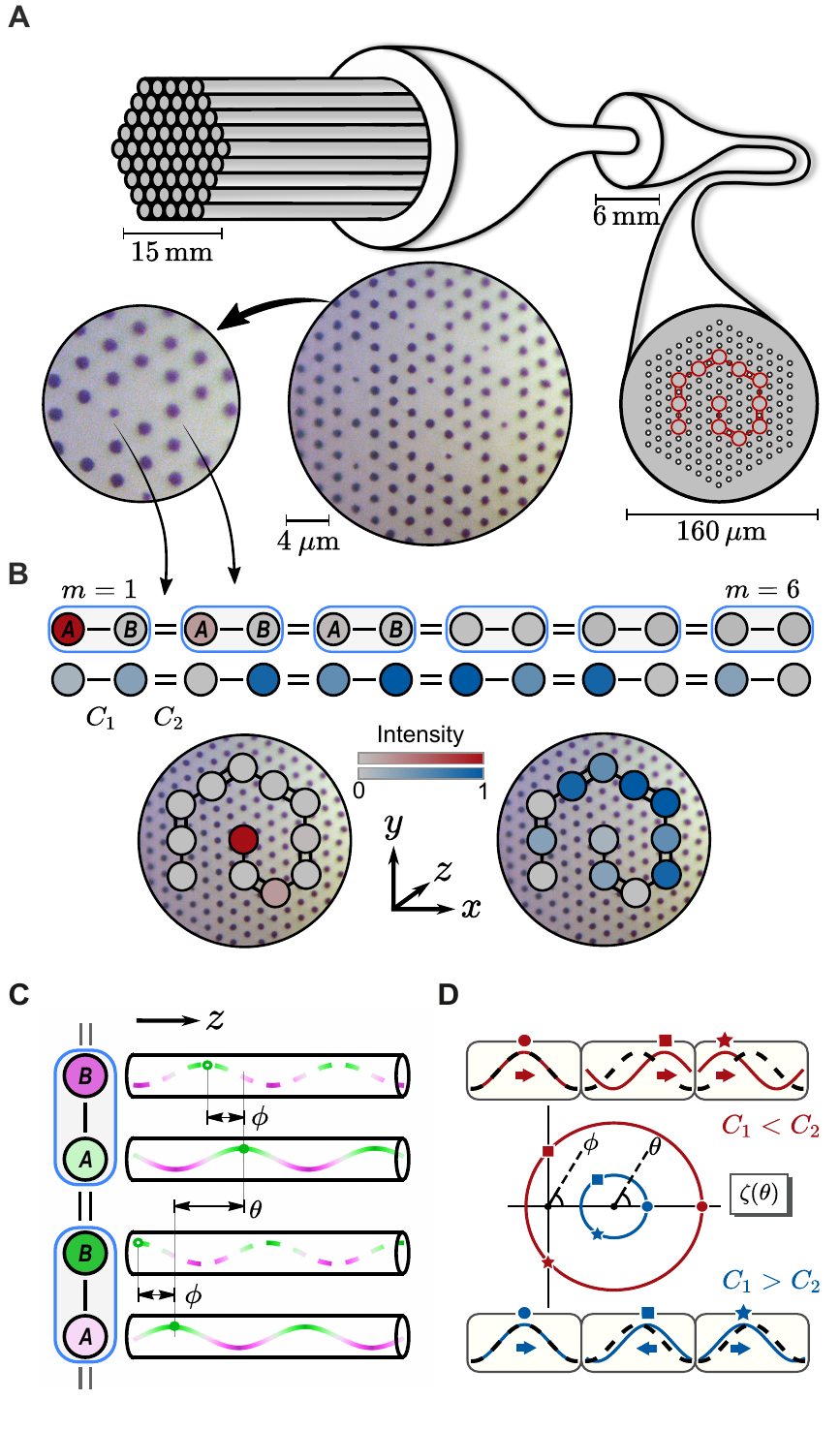}
\caption{\textbf{Topological states in fibre.}  \textbf{A}, Fibre fabrication process: using a fibre drawing tower, stacked glass capillaries are formed into canes and then fibre. Right: the transverse structure is preserved. Middle: optical micrograph of fibre cross-section with dark air holes and lighter grey glass. Left: three glass cores and the different-size air holes between them. \textbf{B}, Smaller air holes correspond to stronger inter-core coupling (double bonds), and larger air holes to weaker coupling (single bonds). A Su-Schrieffer-Heeger (SSH) chain~\cite{Su1979} has alternating couplings and two cores ($A$ and $B$) per unit cell (blue boxes). 
Its supermodes include topological edge states (top, red) and bulk states (middle, blue). 
The chains are wrapped into compact spirals within the fibre cross-section (bottom). \textbf{C}, The topological invariant $\nu$ is encoded in the phase differences ($\phi$ within a unit cell and $\theta$ for neighbouring cells) between optical waves propagating in neighbouring cores (see text).
\textbf{D}, The function $\zeta(\theta)$ has argument $\phi$ and draws a circle in the complex plane. The circle winds around the origin for $C_1 < C_2$ (topological, red), but not for $C_1 > C_2$ (trivial, blue). Boxes: phase difference $\phi$ between $A$ cores (solid) and $B$ cores (dashed) for corresponding $\theta$ values on the circles.
} 
 \label{fig:1}
\end{figure}

We create a topological lattice that propagates light over metre scales governed by its micrometre-scale cross-section by engineering the coupling between cores (Fig.~\ref{fig:1}\textbf{A} and Materials and Methods E). We focus on a canonical model exhibiting a topological invariant, the Su-Schriefer-Heeger (SSH) chain~\cite{Su1979}, defined by identically shaped cores and alternating inter-core coupling strengths. To achieve this, the design alternates between small and large air holes between neighbouring cores. The two ends of the chain include an additional small air hole to create a chain of cores uniform in shape. We curl the chain up inside the fibre to form a spiral (Fig.~\ref{fig:1}\textbf{A}--\textbf{B}) and observe that both ends of the spiral host topologically protected edge states.

Light is guided by fibre in normal modes. Consider a single optical polarisation of the transverse electric field $\mathbf{E}$ at wavelength $\lambda$, which we denote $\psi(x,y)$ ($=E_x$ or $E_y$). The eigenvalue equation for $\psi(x,y)$ and the modal propagation constant $\beta$ (that is, the longitudinal $z$-component of the wavevector) takes a form analogous to the time-independent Schr\"{o}dinger equation. In this analogy, the square of the local refractive index $n(x,y)$ plays the role of the confining potential for light. Explicitly, the eigensystem reads: 
\begin{equation}\label{eq:scalarwaveeq}
    \nabla_{\perp}^{2}\psi(x,y) + 
    k_0^2 n^2 \psi(x,y) = \beta^2 \psi(x,y),
\end{equation}
where $\nabla_{\perp}^{2} \equiv \partial_x^2 + \partial_y^2$ is the transverse Laplacian and $k_0 = 2 \pi/\lambda$ is the free-space wavenumber (see derivation in Materials and Methods A). In our system, high symmetry within each core leads to minimal polarisation coupling, so that the mode profiles are degenerate under rotations of the electromagnetic fields. 
A solution of Eq.~(\ref{eq:scalarwaveeq}) is an electric-field profile $\psi(x,y)$ that propagates unchanged in time $t$ along the longitudinal $z$-direction as $\psi(x,y) e^{i (\omega t - \beta z)}$, with angular frequency $\omega = 2 \pi c/\lambda$ in terms of the speed of light in vacuum, $c$.

The similarity between Eq.~(\ref{eq:scalarwaveeq}) and the Schr\"{o}dinger equation inspires the design of fibre using analogies with electronic band structures. For example, the tight-binding model predicts that for weakly coupled cores, supermodes obey the matrix equation $\mathrm{C} \mathbf{u}=\Delta\beta \, \mathbf{u}$, where $\Delta \beta$ is the change in propagation constant for a supermode relative to the single-core propagation constant, and $\mathbf{u}=(a_1,b_1,a_2,b_2,\dots)$ is a complex vector describing both amplitudes and phases of the single-core modes constituting the supermode. For the SSH chain (Fig.~\ref{fig:1}\textbf{B}), the coupling matrix $\mathrm{C}$ contains just two (nearest-neighbour) elements: we denote $C_1$ to be the coupling strength between cores $A_m$ and $B_m$ within the $m$th unit cell, and $C_2$ the coupling between $B_m$ and $A_{m+1}$ across unit cells.

The topological invariant $\nu$ is encoded in the phase differences between optical waves propagating in neighbouring cores.
When the chain is infinitely long, the supermodes $\mathbf{u}$ can be decomposed into an eigenbasis of supermodes defined by the relations $a_{m+1}=e^{i\theta}a_m$ and $a_m = \pm e^{i\phi}b_m$.
For these normal modes, 
each core contains the same light intensity, with the phase difference $\phi$ between the two sublattices within a single unit cell, and the phase difference $\theta$ between cores on the same sublattice and in neighbouring cells, as illustrated in Fig~\ref{fig:1}\textbf{C}.
In other words, $\theta$ is the reciprocal-space variable along the chain.
The phase differences $\phi$ and $\theta$ are related via the coupling matrix.
In the eigenbasis, $\mathrm{C}(\theta)$ is a $2\times2$ anti-diagonal matrix with entries $\zeta(\theta) = C_1+C_2e^{i\theta}$ (and its complex conjugate, see Materials and Methods A).
The phase $\phi$ is fixed as the complex argument of the function $\zeta(\theta)$, plotted as circles in Fig.~\ref{fig:1}\textbf{D}.
The map $\phi=\arg\zeta(\theta)$ relates two phases and is, therefore, characterised by an integer invariant called the winding number $\nu$.

If the circle $\zeta(\theta)$ circumscribes the origin, then the winding number $\nu$ of the SSH chain is equal to one, which occurs when the intra-cell coupling is weak, $C_1 < C_2$. For strong intra-cell coupling $C_1 > C_2$, this topological invariant is zero. Mathematically, we compute

\begin{equation}\label{eq:windingnr}
    \nu = \frac{1}{2\pi}\int_0^{2\pi}\frac{d\phi}{d\theta}\,d\theta 
    = \begin{cases}
        \ 1 & \text{if }C_1<C_2, \\
        \ 0 & \text{if }C_1>C_2.
    \end{cases}
\end{equation}

Physically, as we change $\theta$ by a full period, we note that for $\nu = 1$, the phase difference $\phi$ between cores $A$ and $B$ undergoes a full period ($2\pi$) shift (Fig.~\ref{fig:1}\textbf{D} top boxes). For $\nu = 0$, $\phi$ stays close to zero (Fig.~\ref{fig:1}\textbf{D} bottom boxes).

Topological band theory predicts that an integer invariant characterising an unbounded system has profound consequences once boundaries are introduced. 
Due to the alternating  coupling coefficients $C_1$ and $C_2$ in the SSH chain, a gap opens up (centred at the single-core propagation constant $\beta$), which has gap size $2 |C_1-C_2|$ (see Fig.~\ref{fig:betavalhist}). 
No bulk modes are allowed to have propagation constants inside this band gap. 
Nevertheless, the band gap can host modes localised at a boundary, and the topological invariant encodes information about these boundary modes.
A heuristic argument explains this connection. An invariant can only change discontinuously and only when the band-gap closes, such as at a boundary between a topological chain and a trivial cladding. This transition necessitates a closing and re-opening of the gap, guaranteeing a mode to exist at the boundary. Because this argument does not rely on the boundary's geometry, the band topology endows this edge mode with protection against disorder. 

\paragraph*{}
\section*{Results and discussion}

\begin{figure*}[tbp]
\centering
\includegraphics[width=\textwidth]{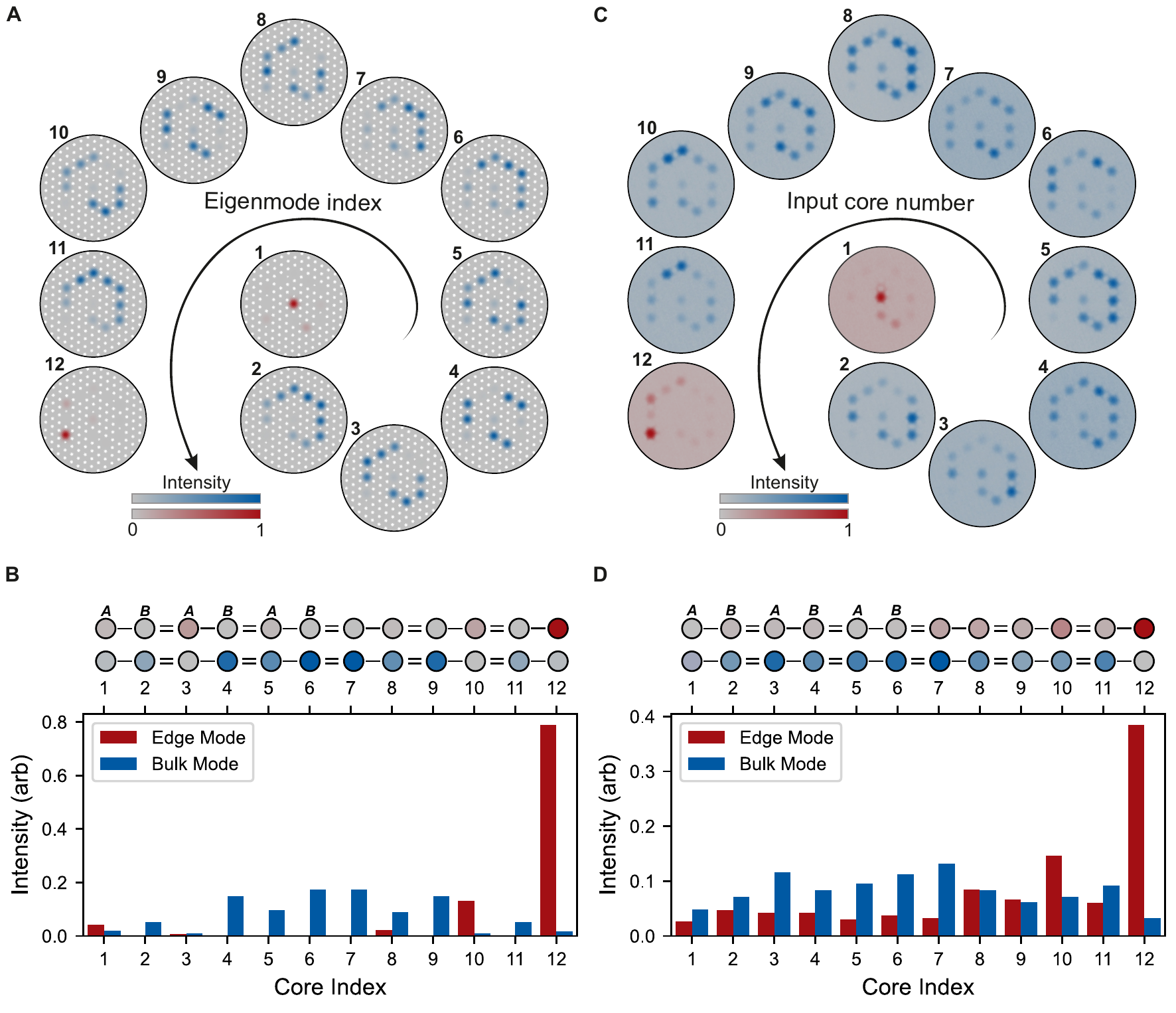}
\caption{\label{fig:2}
\textbf{Bulk and edge states.} \textbf{A}--\textbf{B}, Finite-element simulations 
\textbf{A}, The relative intensity of the electric field for each supermode (doubly degenerate due to polarisation). The two edge states (in red) correspond to supermodes that we label 1 and 12. \textbf{B}, Comparison of the intensity profiles between bulk mode $6$ (blue) and edge mode $12$ (red). \textbf{C}--\textbf{D}, Light propagation experiments: We inject light (wavelength \SI{636}{\nano\metre}) into a core (\emph{input core number}) of a TopoPCF with length \SI[separate-uncertainty = true]{17.0 \pm 0.1 }{\centi\metre} and image the output intensity profiles. For input cores 2 to 11, light excites bulk modes and spreads through the chain. For input cores 1 and 12, light remains localised to the topological edge states. \textbf{D}, Quantitative comparison between the edge mode excited via core 12 and a bulk mode excited via core 4. The topological edge state demonstrates an asymmetry between $A$ and $B$ sublattices characteristic of the SSH chain. }
\end{figure*}

We fabricated a photonic crystal fibre from silica glass by the stack-and-draw process. Our fibre contains 12 cores coupled to form an SSH chain, as illustrated in Fig.~\ref{fig:1}\textbf{A}. Using simulations, we estimate the coupling strengths within the fibre to be: $C_1 = \SI{166}{\per \metre}, C_2 = \SI{352}{\per \metre}$ for light with a wavelength of \SI{700}{\nano \metre}. We verify that our fibre supports chain-edge supermodes through a combination of simulations and experiments shown in Fig.~\ref{fig:2}.

In Fig.~\ref{fig:2}\textbf{A}--\textbf{B}, we numerically solve Maxwell's full vector equations, finding the propagation constants and field profiles of the supermodes supported by our fibre. 
These supermodes can be separated into two classes: bulk states (labelled 2--11) and  exponentially localised edge modes (1 and 12). The bulk states predominately contain contributions from cores that make up the body of the chain, and have propagation constants that lie in the bulk bands, above and below a topological band gap (see Materials and Methods B). 
Two topological edge modes  are found within this band gap. These edge modes exist at $\Delta \beta \approx 0$, are highly localised to the end of the chain, and are robust to disorder smaller than the band gap. 

In experiments (Fig.~\ref{fig:2}\textbf{C}), light is injected separately into each of the twelve cores by butt-coupling from a single-mode PCF. For each input condition, we show the intensity distribution after light has propagated through the fibre. When light couples into cores 2 to 11, it excites a combination of bulk modes. When exciting bulk modes, light couples across all cores in the body of the chain and, after propagation, exits as a distribution of intensities across the bulk cores (blue in Fig.~2\textbf{C}--\textbf{D}). By contrast, light coupled into the two edge cores predominantly excites the topological modes and remains confined to the edge (red in Fig.~2\textbf{C}--\textbf{D}). Notably, the topological edge states exhibit an asymmetry between the $A$ and $B$ sublattices, showing an alternating intensity profile characteristic of the SSH chain. This intensity profile arises due to the interaction between the topological edge modes and the constraints enforced by sublattice symmetry. 

To confirm the topological origin of the edge states in Fig.~\ref{fig:2}, we directly measure the winding number~$\nu$~\cite{Cardano2017detection,Longhi2018probing,Wang2019}. We augment the method from Ref.~\cite{Wang2019}, with the notable difference that instead of changing the propagation length, we scan over the optical wavelength $\lambda$ using the setup in Fig.~\ref{fig:3}\textbf{A} (see Supplementary Videos for the wavelength-dependent output and Materials and Methods F). 
 With light injected into a core in the bulk, we scan the optical wavelength and observe a change in the intensity distribution at the output of the fibre. This change sees light move from cores belonging to the A sublattice to the B sublattice. In the Materials and Methods G, we connect this change in sublattice intensity distribution with the winding number that characterises the system, giving us access to an experimental observable defining the distinct topological invariants.
At the output, the topological invariant $\nu$ can be computed as twice the weighted intensity difference $I_\textrm{d}(\lambda)$, also called the mean chiral displacement~\cite{Cardano2017detection}, between the $A$ and $B$ sublattices averaged over all wavelengths~$\lambda$:
\begin{equation} \label{eq:chiral-disp}
\nu = 2\left\langle I_\textrm{d}(\lambda) \right\rangle_\lambda = 2 \, \Big\langle \sum_m m (I_A-I_B)  \Big\rangle_{\!\lambda},
\end{equation} 
where $I_A=|a_m|^2$ and $I_B=|b_m|^2$ 
are the intensities in each sublattice and the sum runs over all unit cells $m$ (see derivation in Materials and Methods G). 

\begin{figure*}[tbp]
\includegraphics[width=\textwidth]{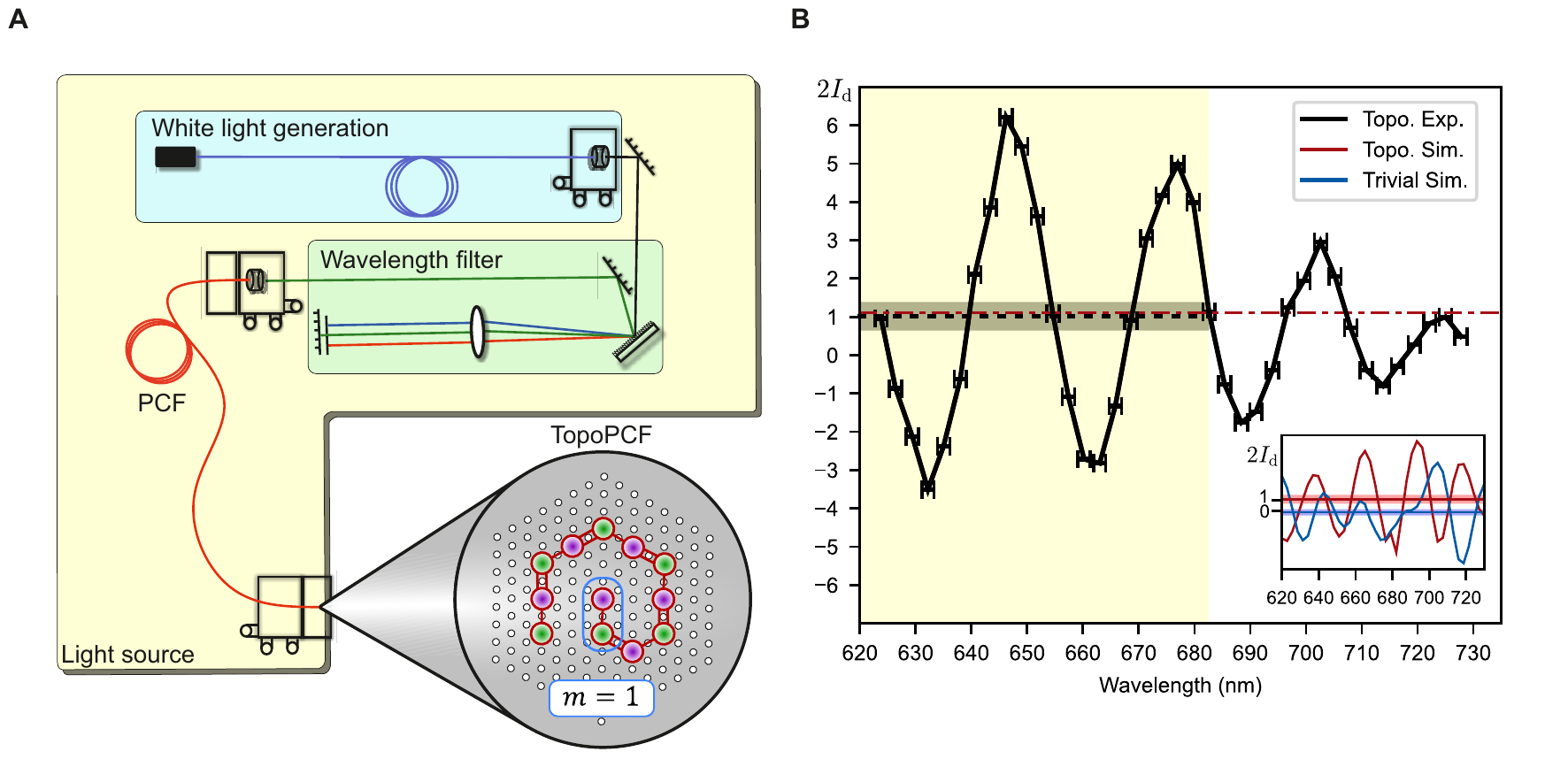}
\caption{\label{fig:3}
\textbf{Direct observation of topological invariant.} 
\textbf{A}, Experimental setup consisting of a variable-wavelength light source coupled into a single TopoPCF core (see Materials and Methods F and Supplementary Videos). \textbf{B}, Real-space measurement of winding number. For light injected into a bulk core, the solid black line is twice the weighted intensity difference between the $A$ and $B$ sublattices, $I_\textrm{d} = \sum_m m (I_{A} - I_{B})$, where $I_{A,B}$ is the normalised sublattice intensity within one unit cell and the sum is over all unit cells, $m = 1\ldots6$. We vary the wavelength over a \SI{104}{\nano\metre} range and propagate through \SI{17}{\centi\metre} of fibre. We calculate the mean of $2 I_\textrm{d}$ to be $1.01\pm0.33$ over the highlighted range to confirm the system's non-trivial winding number (see text). Error bars are set by wavelength filter resolution and the grey confidence interval is the standard deviation of the \emph{unweighted} intensity difference $\sum_m (I_{A} - I_{B})$. Inset: Simulation data for $I_\textrm{d}$ comparing fibre geometries with winding numbers 0 (blue) and 1 (red). For comparison, the topological simulation average is also plotted as a dashed red line in \textbf{B}.}
\end{figure*}

We choose a wavelength range sufficiently broad to capture the characteristic oscillations of $I_\textrm{d}(\lambda)$.
Furthermore, we average $I_\textrm{d}(\lambda)$ over a wavelength sub-range such that the quantity $\left\langle I_\textrm{d}(\lambda) \right\rangle_\lambda$ remains invariant under any renumbering of the unit cells. 
To do so, we pick the wavelength interval over which the measured \emph{unweighted} intensity difference $\sum_m (I_A-I_B)$ vanishes, as explained in Materials and Methods G.
We then use the standard deviation of this unweighted intensity difference as an unambiguous choice for the error. We plot $2{I}_\textrm{d}(\lambda)$ in Fig.~\ref{fig:3}\textbf{B} and measure $2\langle I_\textrm{d}(\lambda)\rangle_\lambda$ to be $1.01\pm0.33$, confirming bulk topology and bulk-boundary correspondence. The simulations agree with our experimental conclusions, and additionally demonstrate that $\langle I_\textrm{d}(\lambda) \rangle_\lambda = 0 $ for the topologically trivial case (see Fig.~\ref{fig:3}\textbf{B} inset).
\onecolumngrid

\twocolumngrid

\begin{figure}[tbp]
\centering
\includegraphics[width=\columnwidth]{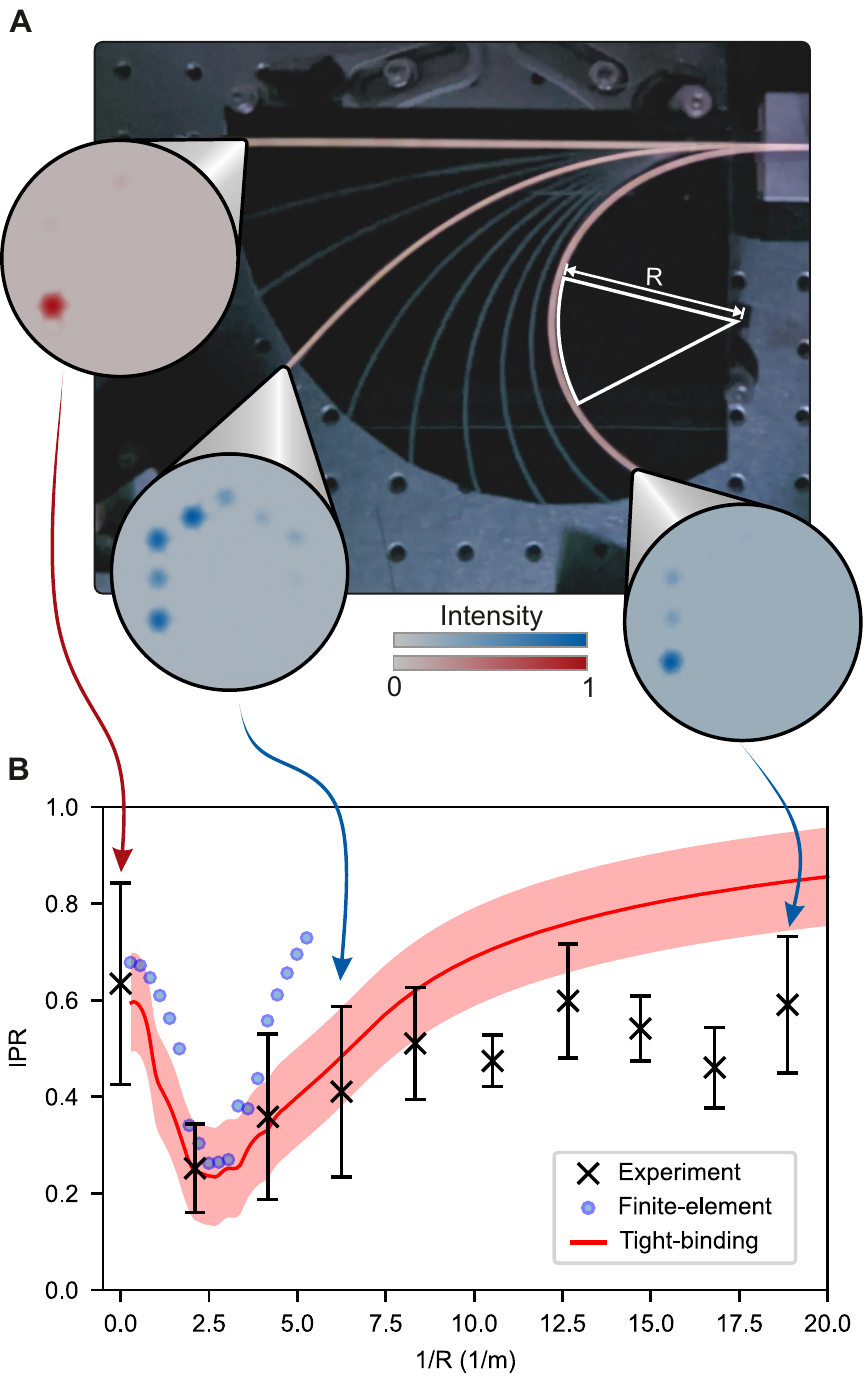}
\caption{\label{fig:4}
\textbf{Switchable topology via bending.} \textbf{A}, Experimental setup for investigating the effects of bending. In the photo, three fibre segments are lit up to show configurations with decreasing radii of curvature $R$. Insets: output intensity after light is injected into edge core 12 (arrows indicate values on the curvature axis in \textbf{B}). As curvature increases, topology is lost and the edge mode delocalises, before trivially relocalising due to bending-induced disorder. \textbf{B}, Inverse participation ratio [$\textrm{IPR}=(\sum_k I_k^2)/ (\sum_k I_k)^2$, where the sum is over all core intensities $I_k$] quantitatively showing this non-monotonic behaviour (see Materials and Methods D). First, the topological edge state is lost (decreasing IPR) before disorder-induced localisation sets in at higher curvature $1/R$ (increasing IPR). Experiment (crosses): six independent measurements per data point (error bars: standard deviation). Finite-element simulation (circles): a single realisation of a bent fibre. Tight-binding model (red line): mean IPR for eigenmodes in 40,000 realisations of on-site disorder for the SSH chain  (shaded interval: standard deviation). 
}
\end{figure}

  Bulk-boundary correspondence and the localised edge modes are a consequence of the topological invariant characterising the system. This invariant must always exist, provided the chain displays sublattice symmetry and experiences sufficiently small disorder (that is, smaller than the photonic band gap). In contrast to a topologically trivial system, the edge modes in our fibre always exist provided the invariant remains well defined. Thus, we can think of the edge modes as inheriting the robustness of the topological invariant. This robustness reveals itself when disorder is introduced to the coupling strengths. If the couplings in the chain contain disorder, topologically trivial modes can be perturbed and even destroyed, whereas topological edge states remain immune. For cases that respect sublattice symmetry, the topological mode remains protected as long as the ratio $C_{1}/C_{2}$ is less  than unity. By contrast, if the disorder violates sublattice symmetry, the protection of the edge modes is weaker.

 When on-site disorder is introduced into our chain, the sublattice symmetry is broken. Under these conditions, the topological edge mode loses its guaranteed protection, and can now be destroyed while the ratio $C_{1}/C_{2}$ is less than unity. However in order for the mode to be destroyed, the on-site disorder must be sufficiently large when compared to the size of the band gap~\cite{CHEN2020126168}. In Materials and Methods \textbf{C}, we explore the effects of on-site disorder that exist in our fibre. We find that for realistic structural imperfections, the strength of disorder is less than $3.2\%$ of the photonic band gap. For these small values of the disorder, the edge modes remain identical to the ideal case that respects sublattice symmetry. 

Our fibre platform gives rise to topological phenomena inaccessible in short, rigid waveguides and solid-state devices. One such phenomenon is the re-configurable control of topological states through geometric bending, only accessible due to the unique physical flexibility of fibre (Fig.~4). In the presence of a bend with radius $R$, the propagation constant of each core changes due to this geometric perturbation by $\Delta\beta \approx 2\pi \Delta x / (\lambda R)$ for cores separated by distance $\Delta x$ (see Materials and Methods D and Refs.~\cite{Schermer2007,Marcuse1976}).  This geometric control enables us to globally reconfigure the SSH chain with a single macroscopic bend. Such reconfigurability is in contrast to metamaterial systems where individual unit cells can be adapted, but changing the entire lattice presents a challenge. 

The changes in propagation constant due to fibre bending map directly onto on-site disorder in the system's coupling matrix.
This mapping enables us to analytically predict the response of the topological edge mode to bending. Because on-site disorder does not preserve the chiral symmetry of the SSH chain, the topological invariant is no longer guaranteed. This loss of symmetry destroys the localising topological protection and allows light in the edge mode to spread across the bulk modes. We observe this in the insets in Fig.~\ref{fig:4}A: As the curvature is increased, the topological edge mode becomes delocalised. Upon further bending this trend begins to reverse and the on-site disorder grows sufficiently large to induce Anderson localisation~\cite{Scollon2020,Mondragon-Shem2014}, which reconfines light within the initially excited core. To quantify this localisation, we calculate the inverse participation ratio (IPR) $\textrm{IPR}=(\sum_k I_k^2)/ (\sum_k I_k)^2$, where the sums run over all cores $k = 1$ to $12$. We plot the IPR as a function of fibre curvature in Fig.~\ref{fig:4}. We predict the breakdown of topological protection at the bend radius $R^*$, at which the ratio of disorder $\Delta\beta$ to the average coupling strength $C$ is near one [$\Delta \beta \approx C = (C_1 + C_2)/2$]. As found in Materials and Methods D, $1/R^* \approx 1.28 \lambda C /(2\pi \Delta x) \approx \SI{3}{m^{-1}}$, in agreement with Fig.~\ref{fig:4}B. This bending radius characterises a simple reversible mechanism for globally breaking and restoring topological protection.  

To conclude, we have used the stack-and-draw technique to make hundreds of metres of topological fibre for visible light. We observe the topology and control the corresponding edge states through fibre bending, a mechanism inaccessible to on-chip waveguides.  We envision topological fibre as a flexible platform that allows topological effects to be applied in a scalable way to classical and quantum photonic networks. 

\section*{Materials and Methods}

\renewcommand\thesubsection{\Alph{subsection}}
\subsection{Theory of the photonic SSH chain}

Light propagating in a waveguide is described by Maxwell's equations   in matter with no free charges or currents, constant permeability $\mu\approx\mu_0$, and spatially varying permittivity $\epsilon/\epsilon_0 = n^2(x,y)$:
\begin{equation}\label{eq:Maxwell}
    \nabla\cdot(n^2\mathbf E)=0,\quad 
    \nabla\times\mathbf H = \epsilon_0 n^2 \partial_t\mathbf{E},
\end{equation}
together with $\nabla\cdot\mathbf H = 0$ and $\nabla\times\mathbf E = -\mu_0\partial_t\mathbf{H}$. Using the curl-curl identity and the expansion of the first equation in Eq.~(\ref{eq:Maxwell}), we obtain the vector wave equation
\begin{equation}\label{eq:vectorwaveeq}
    \nabla^2\mathbf{E}-\frac{n^2}{c^2}\partial_t^2\mathbf{E} 
    +2\nabla\left(\mathbf{E}\cdot\frac{\nabla n}{n}\right)
    =0.
\end{equation}
The final term may be ignored in our system, as the variation in the refractive index is negligible everywhere the electric field is non-evanescent (i.e. in the glass). Translation invariance in the $z$-direction along the fibre allows us to write solutions to Eq.~(\ref{eq:vectorwaveeq}) as superpositions of modes of the form $\mathbf{E} = [\psi(x,y) \hat{\mathbf{e}}] e^{i(\omega t - \beta z)}$, with angular frequency $\omega$ (which is the same in glass as in air), propagation constant $\beta$, and a unit vector $\hat{\mathbf{e}}$ pointing perpendicular to the fibre. Substituting these modes into Eq.~(\ref{eq:vectorwaveeq}) gives the scalar wave equation
\begin{equation}\label{eq:scalarwaveeqM}
    \nabla_{\perp}^{2}\psi(x,y) + (k_0^2 n^2 - \beta^2)\psi(x,y) = 0,
\end{equation}
where $\nabla_{\perp}^{2} \equiv \partial_x^2 + \partial_y^2$, which is Eq.~(1) in the Introduction with $k_0=\omega/c$. 

\begin{figure}[tbp]
\centering
\includegraphics[width=\columnwidth]{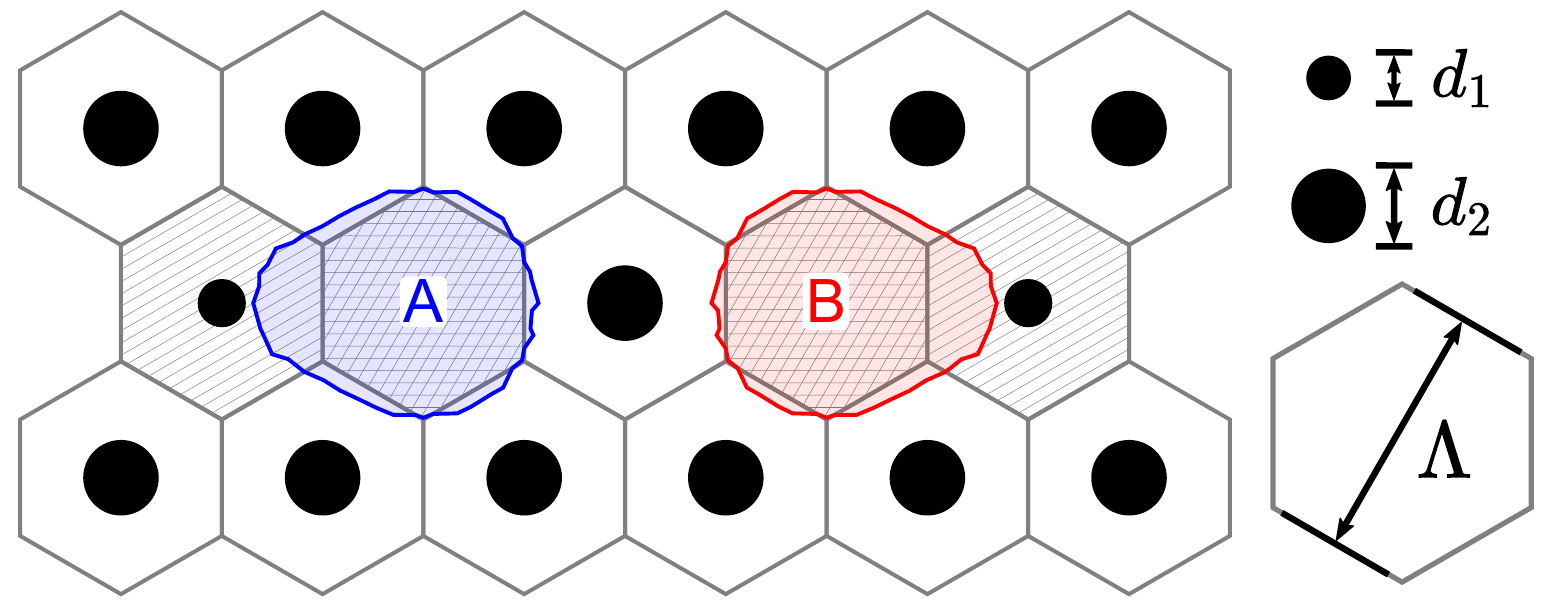}
\caption{\label{fig:methodssketch} \textbf{Discretised multi-core geometry.} The coarse-grained refractive index is $n_g$ on the glass cores (crosshatched) and $n_g-f_j\delta n$ around the small (linearly hatched) and large (unhatched) air holes, where $f_j\propto d_j^2/\Lambda^2$ is the air hole area fraction. A contour line of the single-core mode $\psi_{2m-1}$ localised on an $A$ site is drawn in blue. The contour of $\psi_{2m}$ on the $B$ site is drawn in red. We use exaggerated air hole diameters $d_1 = \SI{0.25}{\micro\metre}$, $d_2 = \SI{0.82}{\micro\metre}$ and pitch $\Lambda=\SI{4.35}{\micro\metre}$.
}
\end{figure}

Let $n_g$ ($n_a$) be the refractive index of the glass (air), and $\delta n=n_g-n_a$ their difference.
In our multicore fibre, the air hole geometry can be discretized with hexagonal platelets to obtain an effective refractive index $n(x,y)$, which evaluates to $n_g$ on the cores, $n_g-f_1\delta n$ around the small air hole flanking the core, and $n_g-f_2\delta n$ around the larger air holes everywhere else. Here $f_k=\frac{\pi}{2\sqrt{3}} d_k^2/\Lambda^2$ is the air hole area fraction in terms of hole diameter $d_k$ and fibre pitch $\Lambda$, see Fig.~\ref{fig:methodssketch}. 

We expand $\psi$ as a supermode $\psi=\sum_j u_j\psi_j$ in terms of single-core modes $\psi_j$. At first approximation, we can solve for $\psi_j$ on the $j$\textsuperscript{th} core in the absence of other cores. That is, we use Eq.~(\ref{eq:scalarwaveeqM}) with a coarse-grained refractive index $n_j(x,y)$ where every $i$\textsuperscript{th} core, with $i\neq j$, is replaced by a large air hole, that is
\begin{equation}
    n_j(x,y) = n(x,y) - \sum\nolimits_{i\neq j}\Delta n_i(x,y),
\end{equation}
where $\Delta n_i$ evaluates to $f_2\delta n$ on the $i$\textsuperscript{th} core and to zero everywhere else. Note that, because the small air hole alternates between being on the left and right side of the core, the single-core mode on an $A$ ($B$) site will be left(right)-heavy, as depicted in  Fig.~\ref{fig:methodssketch}. 

Writing the shift in propagation constant $\Delta\beta$ for a supermode compared to the single-core $\beta$, we subtract the wave equation for the individual cores from the wave equation for the supermode to obtain
\begin{equation}\label{eq:supermodewaveeq}
    \sum_ju_j\left[2\beta\Delta\beta-k_0^2(n^2-n_j^2)\right]\psi_j = 0.
\end{equation}

Multiplying Eq.~(\ref{eq:supermodewaveeq}) on the left by a specific core-mode $\langle\psi_k|$ while approximating  $n+n_j\approx 2n_g$ and  $\beta\approx k_0n_g$, we have
\begin{equation}\label{eq:innerproducts}
    \sum_j \Delta\beta\langle\psi_k|\psi_j\rangle u_j
    = \sum_j\sum_{i\neq j} k_0\langle\psi_k|\Delta n_i|\psi_j\rangle u_j,
\end{equation}
where we use bra-ket vector notation for convenience.
Since the single-core modes are strongly localized, the left-hand side is dominated by $\langle\psi_k|\psi_k\rangle\approx1$, whereas sum on the right-hand side is dominated by $\langle\psi_{j\pm1}|\Delta n_{j\pm1}|\psi_j\rangle\ll1$. 
Hence, for the dominant terms, we must have $\Delta\beta\ll k_0$ and
\begin{equation}\label{eq:couplingmatrix}
    \Delta\beta u_k = \sum_{j=k\pm1} C_{kj} u_j,
\end{equation}
which is our starting point for the tight-binding approach in the Introduction.
These real coupling constants $C_{kj}=k_0\langle\psi_k|\Delta n_k|\psi_j\rangle$ are symmetric since every neighbouring pair of single-core modes are reflections of each other. Furthermore, this alternating left/right-asymmetry of the single-core modes allows us to define the notation $C_1=C_{2m-1,2m}$ and $C_2=C_{2m,2m+1}$ with $C_1<C_2$.

We rewrite Eq.~(\ref{eq:couplingmatrix}) in bra-ket notation as $\textrm{C}|\mathbf{u}\rangle=\Delta\beta|\mathbf{u}\rangle$ and the ket $|\mathbf{u}\rangle=\sum_m a_m|A_m\rangle+b_m|B_m\rangle$ in terms of $a_m=u_{2m-1}$ and $b_m=u_{2m}$. Then, our coupling matrix can be expressed as
\begin{equation}\label{eq:realspacecouplingmat}
     \mathrm{C} = \sum_{m=1}^{M} C_{1} |B_m\rangle\langle A_m|+
        C_{2} |A_{m+1}\rangle\langle B_m| + \mathrm{h.c.}
\end{equation}
Note that the second term introduces coupling between different unit cells (different $m$). To simplify the analysis, we use a discrete Fourier transform (DFT) to switch to a different basis enumerated by discrete angles $\theta_k=2\pi k/M$ for $k=1,\dots,M$:
\begin{equation}\label{eq:DFTbasis}
    |A(\theta_k), B(\theta_k)\rangle = \frac{1}{\sqrt{M}}\sum_{m=1}^M e^{im\theta_k }|A_m, B_m\rangle.
\end{equation}
In this basis, $\textrm{C}$ couples only between states with the same $\theta_k$:
\begin{equation}\label{eq:reciprocalspacecouplingmat}
    \textrm{C}=\sum_{k=1}^M\zeta(\theta_k)|B(\theta_k)\rangle\langle A(\theta_k)| + \mathrm{h.c.},
\end{equation}
where $\zeta(\theta_k)=C_1+C_2 e^{i\theta_k}$. Note that Eq.~(\ref{eq:realspacecouplingmat}) contains coupling to the extraneous core $A_{M+1}$ which the DFT equates with core $A_1$. To remove these periodic boundary conditions, the standard method is to state that the boundary is sufficiently far away by sending $M\to\infty$. 

In the continuum limit, $\theta$ lives on the interval $[0,2\pi]$. We use Eq.~(\ref{eq:reciprocalspacecouplingmat}) to solve the eigenproblem $\mathrm{C} |\mathbf{u}\rangle = \Delta\beta |\mathbf{u}\rangle$ and obtain two solutions for every $\theta$, given by  eigenvectors
\begin{equation}
    |\mathbf{u}_\pm(\theta)\rangle \propto \pm e^{i\phi(\theta)}\,|A(\theta)\rangle +|B(\theta)\rangle,
\end{equation}
where $\phi(\theta)=\arg\zeta(\theta)$, and  eigenvalues
\begin{equation}
    \Delta\beta_\pm(\theta) = \pm|\zeta(\theta)|.
\end{equation}
The two bands of eigenvalues are separated by a gap at $\theta=\pi$ of size $2|C_1-C_2|$ around the single-core propagation constant $\beta$.
Inside this gap no supermodes can propagate through the bulk of the chain.

In terms of the components of $\mathbf{u} = (a_1,b_1,a_2,b_2,\dots)$,  the eigenvectors read
\begin{equation}
    \begin{pmatrix}
    a_m\\b_m
    \end{pmatrix} = \frac{1}{\sqrt{2M}}\begin{pmatrix}
    \pm e^{i\phi(\theta)}\\1
    \end{pmatrix}e^{i\theta},
\end{equation}
which forms the starting point of our definition of topological invariant $\nu$ in the Introduction, as defined in Eq.~2 and illustrated in Fig.~1\textbf{C}--\textbf{D}.

\subsection{COMSOL Simulations} 

We numerically solve the full Maxwell's vector equations defined in Materials and Methods A to find modes of the form $\mathbf{E}(x,y,z) = \mathbf{\tilde{E}}(x,y)e^{i(\omega t-\beta z)}$, including the propagation constants $\beta$ and transverse field profiles $\mathbf{\tilde{E}}(x,y)$ of the normal modes supported by the fibre cross section. We confirm that the coupling to the polarisation of light is sufficiently weak for modes to take the form $\mathbf{E} = [\psi(x,y) \hat{\mathbf{e}}] e^{i(\omega t - \beta z)}$ for arbitrary in-plane polarisation $\hat{\mathbf{e}}$, where we take the time variable $t$ to be fixed. We solve this equation for a finite cross-section selected to match the parameters of the fabricated fibre. Using COMSOL Multiphysics, we run finite-element analysis simulations with varying meshes, materials, and boundary conditions. The COMSOL design process is illustrated in Fig.~S1 in the supplementary materials.

In all our simulations, a bipartite mesh was used, with the core sections featuring a denser mesh than the surrounding cladding. The boundary conditions remain fixed for all simulations: a perfect electric conductor boundary was defined at the perimeter of the cladding. We model the refractive index of silica by implementing in COMSOL a wavelength-dependent Sellmeier equation  with the refractive index of the air holes taking a constant value of 1.

\begin{figure}[tbp]
\centering
\includegraphics[width=\columnwidth]{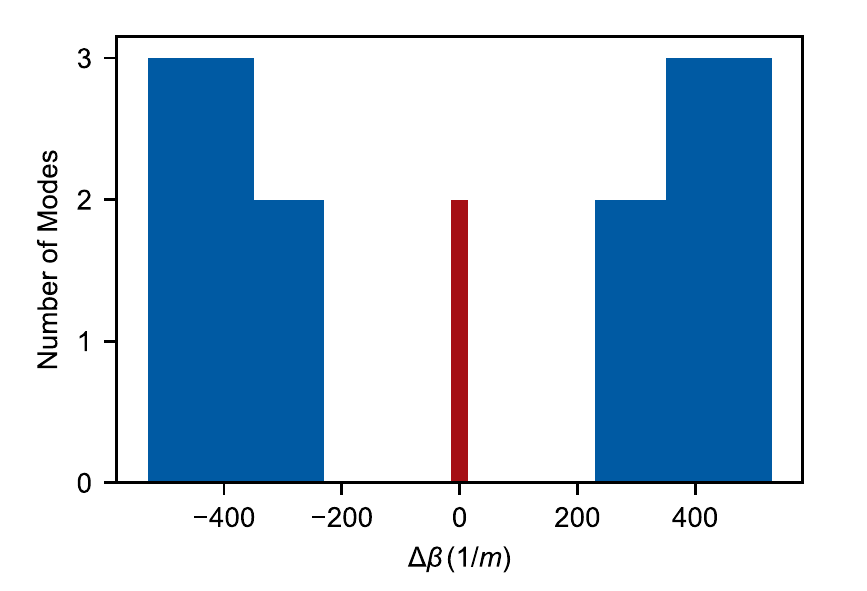}
\caption{\label{fig:betavalhist} \textbf{Propagation constant distribution for simulated supermodes.} A histogram showing the distribution of the propagation constants (for a single polarisation) computed in simulations and plotted in Fig.~2\textbf{A}. The histogram highlights the gap between bulk bands (blue) in the middle of which we find the two topological edge modes (red). As predicted by the analytic theory, these modes exist at $\Delta\beta \approx 0$. 
}
\end{figure}

After specifying materials, mesh sizes, and boundaries, the spectrum of normal modes is calculated. For our TopoPCF, these modes are plotted in Fig.~\ref{fig:betavalhist}. The electric field data from these modes is plotted as $|E(x,y)|^2/\textrm{max}(|E|^2)$ to show relative field intensities and compare to experimental observations. In Fig.~2\textbf{A}, we directly plot the simulated modes, and in Figs.~2\textbf{B},~3\textbf{B},~4\textbf{B} we integrate the intensity across each light-guiding core. We use this integration to normalise the supermode vector $\mathbf{u}$ such that the intensities sum to one.  

\begin{figure*}[tbp]
\includegraphics[width=\textwidth]{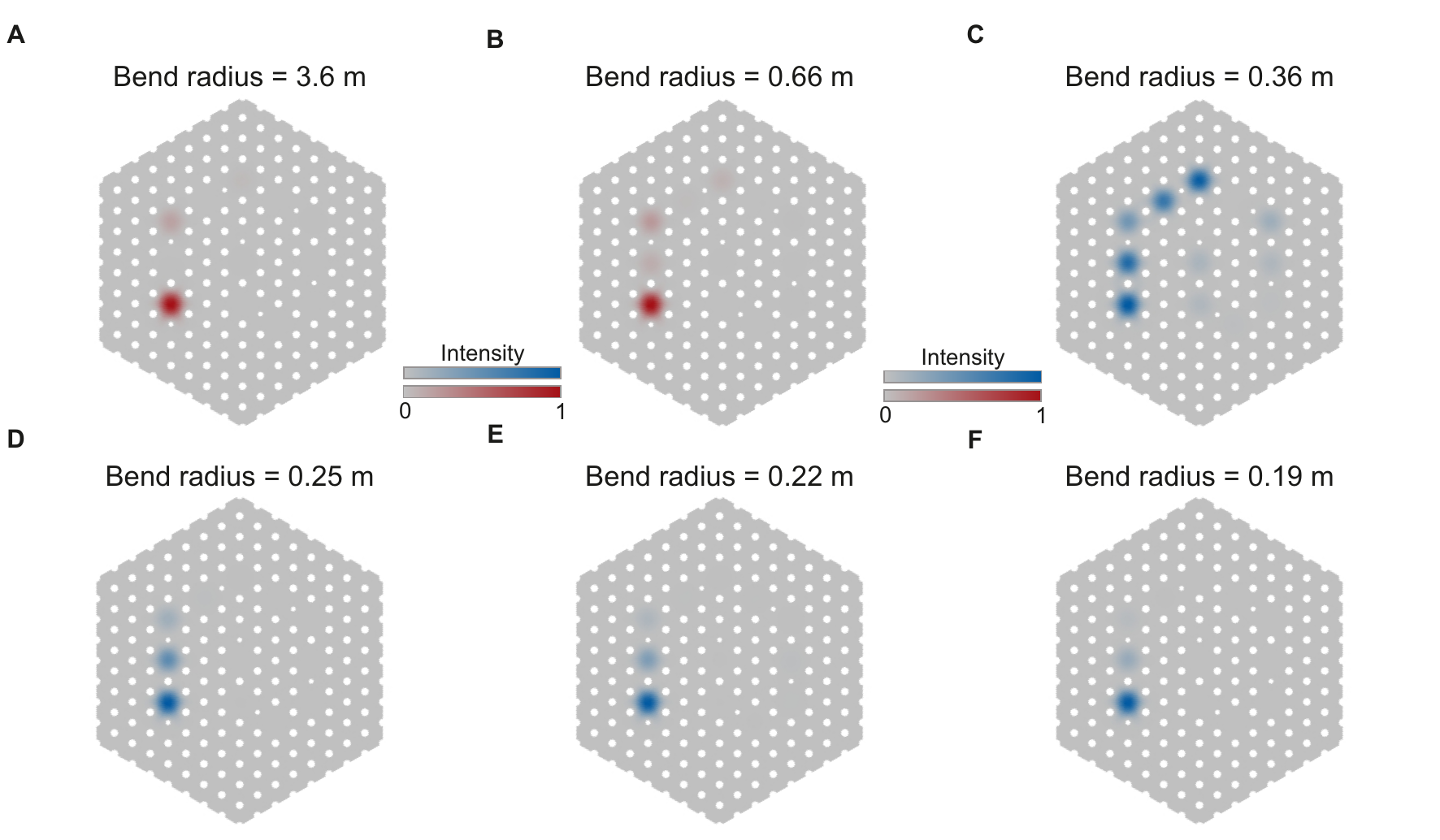}
\caption{\label{fig:bendsims}
\textbf{Bending Simulations}. Results from finite-element simulations of eigenmodes for different bending radii. \textbf{A}, Topological edge mode at large bend radius. \textbf{B}, As we decrease the bend radius, the topological edge mode becomes less localised. \textbf{C}, When the on-site disorder induced by the bend becomes comparable to the coupling strength, the topological protection is broken and the mode becomes delocalised. \textbf{D}, as the bend radius decreases further the mode begins to re-localise due to Anderson localisation. \textbf{E}--\textbf{F}, the mode continues to localise further as the bend radius is decreased.
}
\end{figure*}

In order to simulate fibre bending, the spatial profile $n(x,y)$ of the refractive index of the material was modified, which allows normal modes to be calculated in a bent fibre using our two-dimensional model of the cross section. Following the work in Ref.~\cite{Marcuse1976, Schermer2007}, the refractive index was modified to be $n_{\textrm{bend}} = n_{\textrm{silica}}(1+q/R_{\textrm{eff}})$, where $R_{\textrm{eff}} = 1.28R_{\textrm{bend}}$ and $q$ is the perpendicular distance to the bend axis, to accommodate the change that light would experience in a bent fibre. As highlighted in Ref.~\cite{Schermer2007}, an effective bending radius $R_{\textrm{eff}}$ was used to account for both geometric and stress-optic effects. A selection of the intensity plots found through bending simulations can be found in Fig.~\ref{fig:bendsims}. 

\begin{figure*}[tbp]
    \centering
    \includegraphics[width=\textwidth]{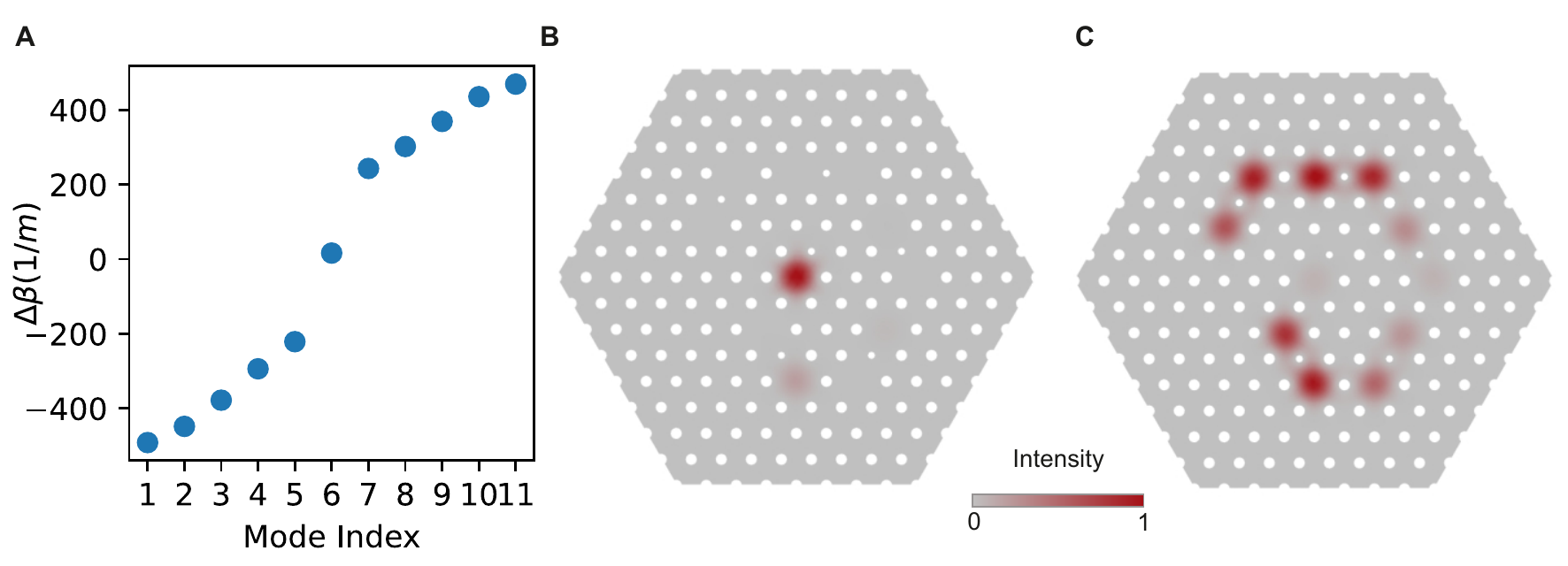}
    \caption{ \textbf{11-core SSH chain.} \textbf{A}, Propagation constants for the 11 supermodes of the shortened chain. The plot highlights the presence of a band gap featuring a single zero mode, localised to just one edge of the chain. \textbf{B}, Intensity plot of the topological edge mode lying in the band gap. This edge mode stays fixed to the core at the centre of the fibre because at that edge of the chain the topological unit cell is not cut. That is, the boundary conditions at the center are the same as in the 12-core fibre that we fabricated, but the boundary conditions at the other end of the chain are different. \textbf{C}, Intensity plot for a bulk mode, which now contains contributions from core 11 at the outer end of the chain. As expected, because the unit cell is cut at this end, this edge no longer hosts a topological state.}
    \label{fig:cut_chain}
\end{figure*}

 The edges of our SSH chain define the system's boundary conditions by determining whether the end of the chain cuts a topological unit cell. For systems with an even number of cores, symmetry necessitates that both ends feature the same boundary conditions. However, if the chain features an odd number of cores, one side will cut a topological unit cell and subsequently lose its edge mode. Once the system features an odd number of cores, there is no way to remove the topological edge (while respecting sublattice symmetry). The topological edge state can simply be moved from one end to the other by adjusting the coupling strengths, and thus unit cell definitions. In Fig.~\ref{fig:cut_chain}, we show through numerical simulations the difference in boundary conditions and the presence of a single edge mode. Fig.~\ref{fig:cut_chain}A shows the propagation constants of an 11 core chain that both starts and ends on a core belonging to the B sublattice. The key feature in Fig.~\ref{fig:cut_chain}A is the single mode lying in the band gap. We plot the intensity distribution of this mode in Fig.~\ref{fig:cut_chain}B to show that only the topological edge mode at the centre of the fibre now remains. Fig.~\ref{fig:cut_chain}C shows that core 11 (the outer end core of the chain) now contributes to bulk modes of the system, and has been shifted out of the band gap. While this change destroys a single topological edge mode, sublattice symmetry changes are needed to destroy the topology everywhere in the chain.

 \subsection{Preserving Sublattice Symmetry}
 
To confirm that our results are not due to localised disorder, we must ensure that all cores are sufficiently similar. Local structural uniformity ensures that all of the light-guiding cores have the same propagation constant. Uniform propagation constants are required in order to make a direct mapping to the tight-binding model. In our system, all cores have the same immediate environment (up to rotational symmetry). However, due to imperfections induced in fabrication and the perturbations each core may experience from other cores in the fibre, this symmetry cannot be guaranteed in practice. That is, chiral and translational symmetries are only approximate in our fibre. The lack of hard symmetry constraints means that theoretically a topological phase transition can occur without the full closing of the band gap. However, the topological invariant that characterises the system remains well defined as long as the band gap is much larger than the variation in on-site potentials~\cite{CHEN2020126168}.

\begin{figure*}[tbp]
    \centering
    \includegraphics[width=\textwidth]{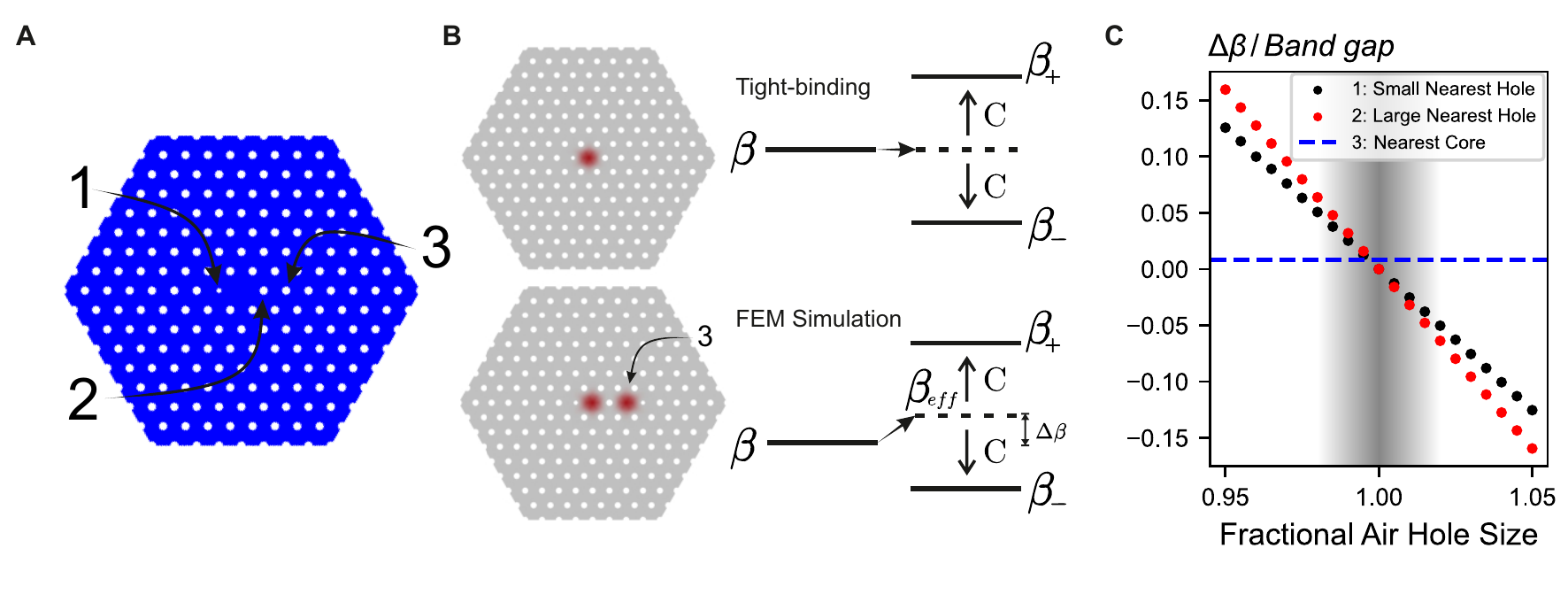}
    \caption{ \textbf{On-site potential changes due to core fluctuations.} \textbf{A}, Cross-section of a single core simulation used to investigate the effects of fabrication defects. Labels 1, 2, and 3 correspond to the different core-changing perturbations that we consider: Label 1 corresponds to the change in radius of the small air hole (mean diameter: \SI{1.0}{\micro\metre}), Label 2 corresponds to the change in radius of the large air hole (mean diameter: \SI{1.6}{\micro\metre}), and Label 3 corresponds to the introduction of a neighbouring core. \textbf{B}, A schematic representation of how the presence of a neighbouring core induces the on-site change $\Delta \beta$ in the propagation constant. This change can be obtained by first calculating the average propagation constant of two supermodes in both the finite element method (FEM) COMSOL simulations and the tight-binding model. The change $\Delta \beta$ is then estimated by taking the difference between these two averages. \textbf{C}, We compare the on-site change induced by the three different types of perturbation (Labeled 1, 2, and 3). The gray region highlights realistic bounds on fabrication disorder, which allows us to estimate the width of the distribution for $\Delta \beta$/(Band gap size) for a realistic fibre, which we find to have a small width of $3.2\%$.} 
    \label{fig:on-site_changes}
\end{figure*}

 We confirm by numerical simulation that on-site disorder is two orders of magnitude smaller than the band-gap size when considering experimental parameters. We consider three changes in geometry that influence the propagation constant of a single core, shown in Fig.~\ref{fig:on-site_changes}. The first two modifications could arise as a result of fabrication imperfections. The two different sizes of air holes immediately adjacent to a core could be over- or under-inflated. This would directly change the propagation constant and hence the effective on-site potential. We model in COMSOL the effects of changes in air hole diameter of up to 5\%, however we typically observe less than $1\%$ change in the geometric parameters of fabricated fibre, as demonstrated in Ref.~\cite{Francis-Jones2016}. We find that with a $1\%$ change to air-hole radius, the core's propagation constant changes by only $3.2\%$ of the band-gap size. 

 In Fig.~\ref{fig:on-site_changes}B, we consider how the propagation constant of a core changes when introducing a new neighbouring core. When a second core is coupled to the first, the band structure splits into two supermodes. In the tight-binding approximation, the cores do not affect each other's on-site potentials and the propagation constants of the supermodes are equally split above and below the single core's propagation constant (see  Fig.~\ref{fig:on-site_changes}B). To estimate the shift in the single-core propagation constant from our COMSOL simulations, we then take the average of the two supermodes. This shift between the averaged supermode propagation constant and the single-core case allows us to estimate the change in on-site potential as a result of the second core. We calculate this shift of the on-site potential to be $0.8\%$ of the system's band-gap size.

\begin{figure*}[tbp]
    \centering
    \includegraphics[width=\textwidth]{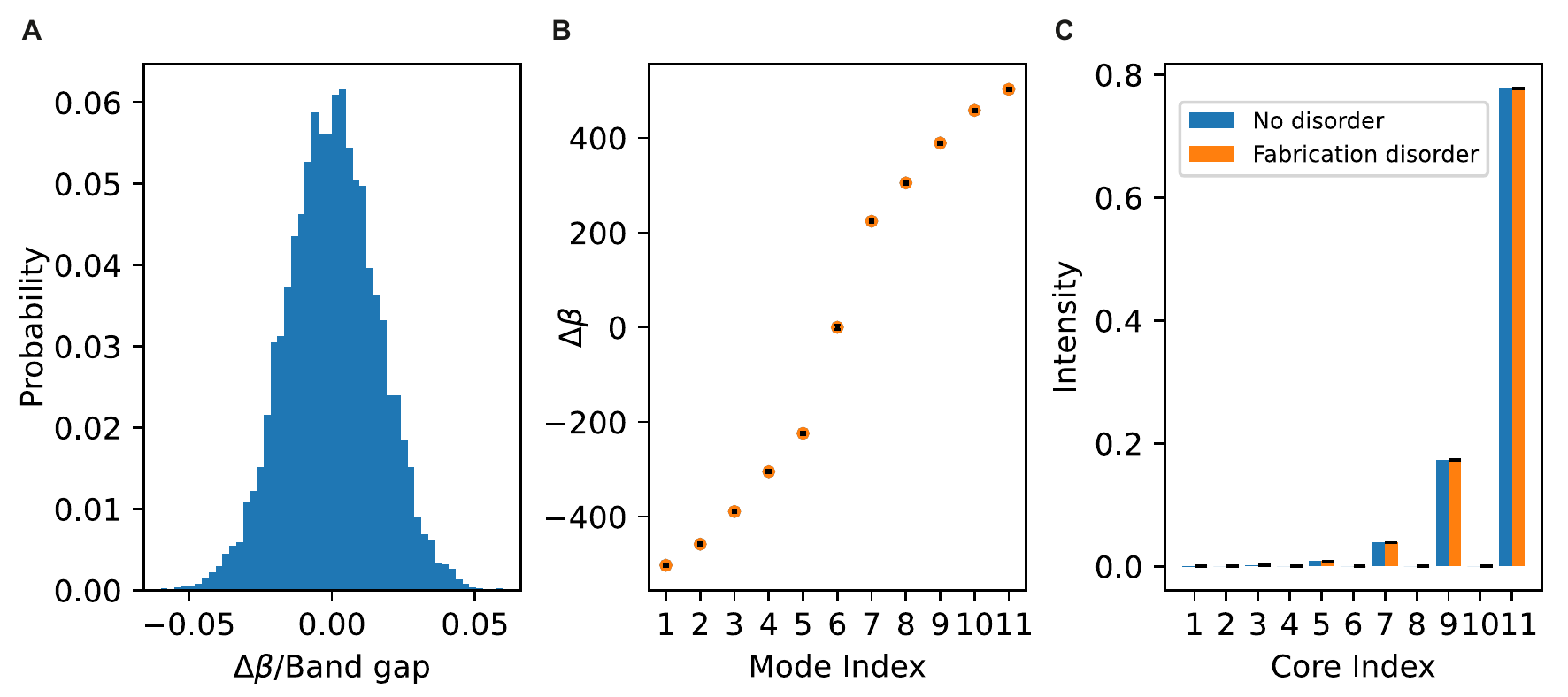}
    \caption{ \textbf{Effects of on-site potential changes.}  \textbf{A}, Using the width estimate defined in Fig.~\ref{fig:on-site_changes}\textbf{C}, we include a normal distribution of on-site disorder in our tight binding model. This graph shows the distribution of on-site disorder for the 10,000 instances for which data is shown in this figure. \textbf{B}, Average propagation constant in the presence of disorder compared to the system with no disorder. The black lines show the standard deviation (smaller than symbol size). We conclude that realistic experimental disorder does not change any of the conclusions of our work. \textbf{C}, Intensity profile for a single edge mode, averaged over the disorder and compared to the case without disorder. The black lines show standard deviation (which is so small compared to the average intensity that it is not visible).}
    \label{fig:on-site_changes_effects}
\end{figure*}

To verify the assumption that the invariant remains protected, we now insert these small changes into the tight-binding model and investigate the resultant mode structure. In Fig.~\ref{fig:on-site_changes_effects}\textbf{A}, we define the distribution from which our random disorder is drawn. This distribution features a standard deviation of 1.6\% change in propagation constant when compared to the band gap. Figure~\ref{fig:on-site_changes_effects}\textbf{B} shows the average propagation constants for the modes of our disordered system, with the black lines showing the standard deviation. We take 10,000 iterations of random on-site disorder from our distribution in Fig.~\ref{fig:on-site_changes_effects}\textbf{A} and apply these  to our tight-binding model. Figure~\ref{fig:on-site_changes_effects}\textbf{C} shows the average edge mode intensity distribution found over the same 10,000 iterations, with the black lines again showing the standard deviation. Figures~\ref{fig:on-site_changes_effects}\textbf{B}--\textbf{C} show no significant deviations from the ideal case without disorder, and still demonstrate a complete topological band gap and a sublattice-polarised edge mode. In Figs.~\ref{fig:on-site_changes_effects}\textbf{B} and \textbf{C}, we use an 11-core SSH chain in place of our 12-core experimental  system. This calculation was chosen because an 11-core chain only has one topological edge mode, at $\Delta \beta = 0$.
\vspace{10pt}
\subsection{Numerical propagation}
Investigating light propagation begins with defining an input vector $\mathbf{u}(z = 0)$, describing the complex amplitude of each individual core mode in the initial excitation. The overlap integral between this input vector and each of the twelve supermodes ($\mathbf{u}_1, \mathbf{u}_2, ..., \mathbf{u}_{12}$) gives the contribution of each supermode present in the initial excitation. By definition, these supermodes propagate along the fibre with only a change in phase (at any fixed time $t$): $\mathbf{u}_1(z) = \mathbf{u}_1(0) e^{-i\beta_1 z}$. By multiplying each supermode by its contribution and phase change, we find the final description of the supermode at a distance $z$ from the input location. After finding the final vectors describing each supermode at the desired distance, we sum these contributions to find the total field distribution in the supermode basis. To return to the core basis, we then perform a second set of overlap integrals for this final supermode vector with each individual core vector, returning a vector containing the contribution of each core to the final state. Multiplying this final state by its complex conjugate returns a vector $\mathbf I(z)= (|a_1|^{2},|b_1|^{2},|a_2|^{2},|b_2|^{2},\dots)$ of relative intensities, describing how much light is in each core in the chain at distance $z$. 

In addition to plotting the propagation of light, the benefit of this numerical propagation is that it allows simulation of the wavelength dependence of the sublattice intensity difference $I_\textrm{d}(\lambda)$ used in Eq.~(3). By finding the supermodes for wavelengths in the range shown in Fig.~3\textbf{B}, numerical propagation can be used to model the output of the fibre after propagating through a given distance $z$. This semi-analytical method allowed us to make use of the detailed simulation results without computationally expensive 3D models, while giving more accurate predictions than a purely analytical approach based on the tight-binding model. 

Following from Materials and Methods B, the change in propagation constant  is: 
\begin{equation}
    \Delta \beta = 2 \pi \Delta x/( R_\textrm{eff} \lambda) = 2\pi \Delta x/( 1.28 R_\textrm{bend} \lambda).
\end{equation}
We modify the coupling matrix $\mathrm{C}$ using diagonal elements corresponding to $\Delta \beta$ for each core. In general, each core's $\Delta \beta$ will have a different value, which in addition depends on the bending axis, therefore playing the role of (diagonal) on-site disorder. In order to separate the general effect of bending from the specific geometry of our design, we now bend this fibre over a range of different bending directions and average the results.

We consider 40,000 different bending directions for each point on the $x$-axis and find the eigenvectors for every system. The inverse participation ratio (IPR) values for these eigenvectors are plotted as a 2D histogram in Fig.~S2 in the supplementary materials. Here we see that at low curvature there are two populations of modes with different IPRs: the localised topological edge states with IPR  $\approx 0.6$ and the bulk states with IPR $\approx 0.15$. As we are interested in the breakdown of topologically protected states we now discard the bulk states and focus on the effects displayed by the edge states. We select the two eigenmodes corresponding to topological edge modes (in every bend realisation) and average their IPRs over all 40,000 points at each step, leaving us with the average effect of a bend on the IPR of a topological edge mode. Averaging the variances of each set of 40,000 bends and then taking the square root allows us to fit an average standard deviation for the entire distribution, as shown in the curve labelled \emph{Tight-binding} in Fig.~3\textbf{B}.

To estimate the point at which the topological edge mode delocalises, we consider the case when the on-site disorder becomes equal to the average coupling strength ($\Delta \beta \approx C = (C_1 + C_2)/2$). At wavelength $\lambda = \SI{500}{\nano\metre}$, the critical curvature corresponding to this disorder is found to be: $1/R^{*}_\textrm{bend} \approx 1.28 \lambda C /(2\pi \Delta x) \approx \SI{3}{m^{-1}}$, which is the value quoted in the `Results and discussion' section.

\subsection{Fibre Fabrication}
The TopoPCF shown in Fig.~1\textbf{A} was fabricated using the stack-and-draw process. The first step was to create a macroscopic preform consisting of silica rods and tubes, stacked by hand in the geometry corresponding to the fibre design. As the design required two air-hole sizes, we used silica tubes with two different ratios of outer-to-inner diameter. The precursor to the larger air holes in the cladding was a tube with \SI{25}{\milli\metre}/\SI{10}{\milli\metre} outer/inner diameter, whereas the smaller air holes, found in between every other pair of cores, were created from a \SI{10}{\milli\metre}/\SI{3}{\milli\metre} tube. The cores were formed from a solid silica rod. Each element was drawn to a diameter of \SI{1.05}{\milli\metre} and stacked into a regular triangular array with the desired layout.

The completed stack was inserted into a \SI{25}{\milli\metre}/\SI{17}{\milli\metre} silica tube and drawn into \SI{4}{\milli\metre} diameter canes, containing the hole pattern required for the fibre. Fibre preforms were created by jacketing canes in \SI{6}{\milli\metre}/\SI{4}{\milli\metre} tubes with brass fittings to enable pressure to be applied to the holes in the cane to prevent collapse during the fibre draw.

The fibre was drawn to a diameter of \SI{162}{\micro\metre} with a high-index UV-cured polymer coating applied during the draw to prevent damage. The PCF used in this work had a pitch  (the centre-to-centre distance for two neighbouring air holes) of approximately \SI{4.4}{\micro\metre}, with a large air hole diameter of approximately \SI{1.6}{\micro\metre}, and small air hole diameter of approximately \SI{1.0}{\micro\metre}. 

\subsection{Characterisation}
Fig.~3\textbf{A} highlights the key features of our experimental setup. We used a \SI{1064}{\nano\metre} sub-nanosecond microchip laser to generate a broadband supercontinuum spanning the visible and near infra-red in a length of single-core PCF. This provided white light in a single spatial mode from which a tunable wavelength range was selected by the monochromator shown in the green box in Fig.~3\textbf{A}. The monochromator was formed of a folded 4-f line containing a reflective diffraction grating and a mechanically adjustable slit to pass the desired wavelength range with a bandwidth of approximately $\SI{2}{nm}$ in the spectral region of interest. The passband was collected by a length of endlessly single mode (ESM) PCF for delivery to the TopoPCF. The pitch of the ESM PCF was approximately \SI{2}{\micro\metre} -- less than half that of the TopoPCF -- enabling the ESM PCF to selectively address individual cores in the TopoPCF by  butt-coupling. After propagation, a magnified near-field image of the output of the TopoPCF was formed on the camera sensor by an aspheric lens. The ESM PCF was positioned with a 3-axis flexure translation stage to couple light into each core of the TopoPCF sequentially in order to obtain output intensity distributions for all bulk and edge excitations shown in Fig.~2\textbf{C}. The total intensity at the output of each core was calculated from the images using Python.

To make the direct observation of the topological invariant shown in Fig.~3\textbf{B}, we first established a wavelength range over which the bandwidth of the input light remained approximately constant. We tuned the passband of the monochromator over a \SI{104}{\nano\metre} range between \SI{632.8}{\nano\metre} and \SI{727.8}{\nano\metre}, and using an optical spectrum analyser we observed that  the bandwidth of the transmitted light did not change significantly. This process also enabled the slit position of the monochromator to be calibrated with transmission wavelength. After calibration the light was coupled from the ESM PCF into core 2 in the TopoPCF. Small wavelength steps (average step size \SI{2.7}{\nano\metre}) were made by varying the slit position in the monochromator without changing any other components of the system. At each wavelength step the output near-field image from the TopoPCF was recorded. From these images the total intensity difference in each unit cell was found.

The bend measurements required additional components in the optical setup. An acrylic board was laser-etched to create circular grooves with $10$ different radii, corresponding to the curvatures seen in Fig.~4\textbf{B}. The board was then fixed in place between the input fibre and the camera in order to hold the TopoPCF at specific bends. After coupling light into core 12, the input was held constant while the fibre was moved between grooves in the board. For each curvature, the output end of the TopoPCF was imaged onto the camera. This process was repeated three times for two different rotational orientations of fibre in order to minimise any specific geometric effects from the axis of the bend. The inverse participation ratio (IPR) was calculated from the total intensity $I_k$ in each core, where $k = 1\ldots12$, using $\textrm{IPR} = (\sum_k I_k^2)/ (\sum_k I_k)^2$. The results are plotted in Fig.~4\textbf{B}.

\begin{widetext}
\begin{center}
\end{center}
\end{widetext}

\setcounter{equation}{0}
\setcounter{section}{0}
\makeatletter
\renewcommand{\theequation}{S\arabic{equation}}
\renewcommand{\thefigure}{S\arabic{figure}}

\setcounter{figure}{0}
\begin{widetext}
\subsection{Weighted intensity difference and the winding number}
Here we derive Eq.~(\ref{eq:chiral-disp}), which allows us to compute the winding number from propagation measurements. We then justify our analysis for this experimental winding-number computation, including the error estimate.

We consider how the intensity distribution changes as light propagates through the system, from being input into a single $A$ or $B$ core in a unit cell labelled by the integer $m^\star$. We define two operators for the \emph{weighted} intensity difference $\mathrm{I}_\textrm{d}$ and the \emph{unweighted} intensity difference $\widetilde{\mathrm{I}}_\textrm{d}$ as
\begin{equation}
    \mathrm{I}_\textrm{d} = \sum_{m} m(\mathrm{I}_{Am} - \mathrm{I}_{Bm})
    \quad\text{and}\quad
    \widetilde{\mathrm{I}}_\textrm{d} = \sum_{m} (\mathrm{I}_{Am} - \mathrm{I}_{Bm}),
\end{equation}
where $\mathrm{I}_{Am}=|A_m\rangle\langle A_m|$ is the intensity of light in the $A$ site (and $\mathrm{I}_{Bm}=|B_m\rangle\langle B_m|$ the intensity in the $B$ site) of the $m$\textsuperscript{th} unit cell and $m$ runs from 1 to 6 in our fibre. The evolution of these quantities along the fibre can be described in terms of the coupling matrix $\textrm{C}$ multiplied by the propagation distance $z$ to yield the expectation value $I_\textrm{d}(z)$:
\begin{align}
    I_\textrm{d}(z) = \langle \bm u(z)|\,\textrm{I}_{\textrm{d}} \,|\bm u(z) \rangle,
    \quad\text{where}\quad
    |\bm u(z)\rangle = e^{-i \textrm{C} z}|\bm u(0)\rangle
    .
\end{align}
We take a Fourier transform along the chain as if the chain were infinitely long, as described in SI A. Identifying $\langle A(\theta)|=(1,0)$ and $\langle B(\theta)|=(0,1)$ in Fourier space, we replace $m\to i\partial_\theta$ and $\textrm{I}_{Am}-\textrm{I}_{Bm}\to\sigma_z$. We write the vector $\bm\zeta = (\mathrm{Re}\zeta,\mathrm{Im}\zeta, 0)$ using complex notation, such that $\textrm{C}=\bm\zeta\cdot\bm{\sigma}$. Here, $\bm{\sigma}=(\sigma_x,\sigma_y,\sigma_z)$ is the vector of Pauli matrices,
\begin{equation}
    \sigma_x =\begin{pmatrix} 0&1\\1&0 \end{pmatrix},\quad
    \sigma_y = \begin{pmatrix} 0&i\\-i&0 \end{pmatrix},\quad
    \sigma_z = \begin{pmatrix} 1&0\\0&-1 \end{pmatrix}.
\end{equation}
We note that our definition of $I_\textrm{d}(z)$ depends on the labelling of the input cell $m^\star$. Supposing, for notational simplicity, that the light starts out in the $B$ site of this unit cell, i.e., $|\bm u(0)\rangle = |B_{m^\star}\rangle$, we have
\begin{align}
    I_\textrm{d}(z)&= \frac{1}{2\pi} \int_{0}^{2\pi} 
        \begin{pmatrix} 0&1 \end{pmatrix}\left(
        e^{+im^\star\theta}e^{+i(\bm\zeta\cdot\bm{\sigma})z}
        \,i\partial_{\theta}\sigma_{z}\,     
        e^{-i(\bm\zeta\cdot\bm{\sigma})z}e^{-im^\star\theta}\,
        \right)\begin{pmatrix} 0\\1 \end{pmatrix}\,
    d\theta
    \\&= \frac{1}{2\pi} \int_{0}^{2\pi} 
        \begin{pmatrix} 0&1 \end{pmatrix}\left(
        e^{i(\bm\zeta\cdot\bm{\sigma})z}\,
        (i\partial_{\theta}+m^\star)\sigma_{z}\,     
        e^{-i(\bm\zeta\cdot\bm{\sigma})z}
        \right)\begin{pmatrix} 0\\1 \end{pmatrix}\,
    d\theta
    \\&= I_\textrm{d}^\star(z) +m^\star\widetilde{I}_\textrm{d}(z).\phantom{\int_0^{2\pi}}
\label{eq:centralizedWID}
\end{align}
Here, $\mathrm{I}_\textrm{d}^\star= \sum_{m} (m-m^\star)(\mathrm{I}_{Am} - \mathrm{I}_{Bm})$ is the \emph{centralised} weighted intensity difference. Note that $\mathrm{I}_\textrm{d}=\mathrm{I}_\textrm{d}^\star$ if the input cell is labelled $m^\star=0$. 

We first show that the winding number $\nu$ can be computed using the wavelength average of
$I_\textrm{d}^\star(z)$. To start, the integrand of $I_\textrm{d}^\star(z)$ can be simplified, by first expanding the exponent of the Pauli vector
\begin{equation}
    e^{\pm i(\bm\zeta\cdot\bm{\sigma})z} = \cos(|\zeta|z)\mathbbm{1} \pm i\sin(|\zeta|z)(\bmh n\cdot\bm\sigma),
\end{equation}
where $\bmh n = \bm\zeta/\Vert\bm\zeta\Vert=(\cos\phi,\sin\phi,0)$ in terms of the phase $\phi=\arg\zeta(\theta)$. 
Using the product of Pauli vectors $(\bm w\cdot\bm\sigma)(\bm v\cdot\bm\sigma) = (\bm w\cdot\bm v)\mathbbm{1}+i(\bm w\times\bm v)\cdot\bm\sigma$ and
the commutation relation $(\bmh n\cdot\bm\sigma)\sigma_z = -\sigma_z(\bmh n\cdot\bm\sigma)$ because $\bm\zeta$ lies in the $xy$-plane, the operator in the $I_\textrm{d}^\star(z)$ integrand reduces to
\begin{align}
    e^{i(\bm\zeta\cdot\bm{\sigma})z}\,
        i\partial_{\theta} \sigma_{z}\,     
        e^{-i(\bm\zeta\cdot\bm{\sigma})z}
    &= i\sigma_z
    \big(c\mathbbm{1} - i s(\bmh n\cdot\bm\sigma)\big)
    \left[\frac{dc}{d\theta}\mathbbm{1} - i \frac{ds}{d\theta}(\bmh n\cdot\bm\sigma)-is\left(\frac{d\bmh n}{d\theta}\cdot\bm\sigma\right)\right]
\\[\jot]
    &= i\sigma_z\left[
        \frac{d}{d\theta}\bigg(\tfrac{1}{2}(c^2-s^2)\mathbbm{1}-ics(\bmh n\cdot\bm\sigma)\bigg)- 
        is^2\left(\bmh n\times\frac{d\bmh n}{d\theta}\right)\cdot\bm\sigma
        \right],
\end{align}
where, for compactness, we abbreviated $c=\cos(|\zeta|z)$ and $s=\sin(|\zeta|z)$.
When integrated against $\theta$, the total derivative of a single-valued and periodic function vanishes by the fundamental theorem of calculus. For the remaining term, we note that $(\bmh n\times\partial_\theta\bmh n)\cdot\bm\sigma=(\partial_\theta\phi)\sigma_z$, leaving us with
\begin{equation}
    I_\textrm{d}^\star(z) = \frac{1}{2\pi} \int_{0}^{2\pi} \sin^2(|\zeta|z)\frac{d\phi}{d\theta}\,d\theta.
\end{equation}

In our experiment, we probe $\lambda$ in the range of \SIrange[range-phrase=--,range-units=single]{600}{750}{nm}. From simulation data, we know that over this range the coupling constants $C_1$ and $C_2$ are well approximated by a linear dependence on $\lambda$, and the coupling ratio $C_1/C_2$ varies on the order of \SI{1}{\percent} per \SI{100}{nm}. As a direct consequence, we can take, for every $\theta$, the direction $\bmh n$ as constant and the magnitude $|\zeta|$ as linear with respect to $\lambda$, with a slope between $|\partial_\lambda C_1 - \partial_\lambda C_2|=\SI{31}{m^{-1}}$ and $|\partial_\lambda C_1+\partial_\lambda C_2|=\SI{92}{m^{-1}}$ per \SI{100}{nm}. Hence, if we take measurements at a length $z>\pi/\SI{31}{m^{-1}} \approx \SI{10}{cm}$, the quantity $|\zeta(\lambda)|z$  samples points across the entire domain $[0,\pi]$ of the $\sin^2$, over which $\sin^2$ has a mean value of $1/2$. In our setup $z=\SI{17}{cm}$, so we can average
\begin{equation}
    \langle I_\textrm{d}^\star(z)\rangle_\lambda 
    = \frac{1}{2\pi} \int_{0}^{2\pi} 
        \left\langle\sin^2(|\zeta|z)\right\rangle_\lambda\,\frac{d\phi}{d\theta}\,d\theta
    =\frac{1}{2\pi} \int_{0}^{2\pi} \frac{1}{2}\frac{d\phi}{d\theta}\,d\theta
    =\frac{\nu}{2}.
\end{equation}
From Eq.~(\ref{eq:centralizedWID}), we see that the average of our uncentralised weighted intensity difference is shifted from the winding number by a factor that depends explicitly on the labelling of the input core:
\begin{equation}
    \langle I_\textrm{d}(z)\rangle_\lambda 
    = \frac{\nu}{2} + m^\star\langle\widetilde{I}_\textrm{d}(z)\rangle_\lambda .
\end{equation}
When averaged over a very large range of wavelengths $\lambda$, this gauge freedom is fixed by the fact that $\widetilde{I}_\textrm{d}(z)$ averages to zero:
\begin{equation}
    \langle \widetilde{I}_\textrm{d}(z)\rangle_\lambda 
    = \frac{1}{2\pi} \int_{0}^{2\pi} 
        \begin{pmatrix} 0&1 \end{pmatrix}\left(\,
        \sigma_{z}\,     
        \left\langle e^{-2i(\bmh n\cdot\bm{\sigma})|\zeta|z}
        \right\rangle_\lambda\,
        \right)\begin{pmatrix} 0\\1 \end{pmatrix}\,
    d\theta
    =0,
\end{equation}
because $e^{2ix}$ is centered at zero on the $[0,\pi]$ domain. In our experiment, we pick a wavelength interval over which $\langle \widetilde{I}_\textrm{d}(z)\rangle_\lambda =0$ so that $\langle I_\textrm{d} (z)\rangle_\lambda$ remains invariant under relabelling. In both approaches, we conclude that $2\langle I_\textrm{d}(z)\rangle_\lambda = \nu$, which is Eq.~(\ref{eq:chiral-disp}) in the `Results and discussion' section. However, the mean of the square of $I_\textrm{d}(z)$ does depend on the labelling $m^\star$. To estimate errors in our calculation of the winding number $\nu$ in an invariant way, we use the standard deviation of the unweighted intensity difference $\widetilde{I}_\textrm{d}(z)$, as plotted in  Fig.~\ref{fig:3}\textbf{b}. This estimate for the error does not depend on the numbering of the unit cells. The error estimate measures the amplitude of the characteristic oscillations of the intensity difference as the wavelength is varied.

\end{widetext}

\section*{Acknowledgments}
The authors thank Marcin Mucha-Kruczynski, Dmitry Skryabin, Jack Binysh, and Marco Liscidini for inspiring conversations.

\subsection*{Funding}
A.S.~acknowledges the support of the Engineering and Physical Sciences Research Council (EPSRC) through New Investigator Award No.~EP/T000961/1 and of the Royal Society under grant No.~RGS/R2/202135. P.J.M.~and J.N.~are supported by the UK Hub in Quantum Computing and Simulation, part of the UK National Quantum Technologies Programme with funding from UKRI EPSRC Grant EP/T001062/1. This material is based upon work supported by the Air Force Office of Scientific Research under award
number FA8655-22-1-7028.

 \subsection*{Author Contributions}
 N.R., J.N., P.M., and A.S.~conceptualized the research. N.R., P.M., and A.S.~designed the fibre. N.R. and P.M.~fabricated the fibre and performed experiments. N.R., G.B., and A.S.~developed the analytic theory. N.R.~performed the simulations. All authors wrote the paper, revised the manuscript critically, and have given approval to the final version. 
 \subsection*{Data Availability}
 All data needed to evaluate the conclusions in the paper are present in the paper and/or the Supplementary Materials. Raw experimental data and corresponding simulation data are available on Zenodo at DOI:10.5281/zenodo.7085818 under an MIT license. 
 
\subsection*{Competing Interests}
 The authors declare that they have no competing interests.

%

%
\section*{Supplementary Information: Topological supermodes in photonic crystal fibre}

\subsection*{Supplementary video captions}
\textbf{Supplementary Video 1: Wavelength dependence of topological edge mode.} We vary the input wavelength from approximately 600nm to 500nm as we couple light into the central core and observe the output.

\noindent \textbf{Supplementary Video 2: Wavelength dependence of topological bulk.} We vary the input wavelength from approximately 600nm to 650nm as we couple light into core 6 and observe the output. Varying wavelength for input into a bulk core allows us to directly compute the topological invariant using Eq.~(3).

\subsection*{Supplementary Figures}
\begin{figure*}[tbp]
\centering
\includegraphics[width=\textwidth]{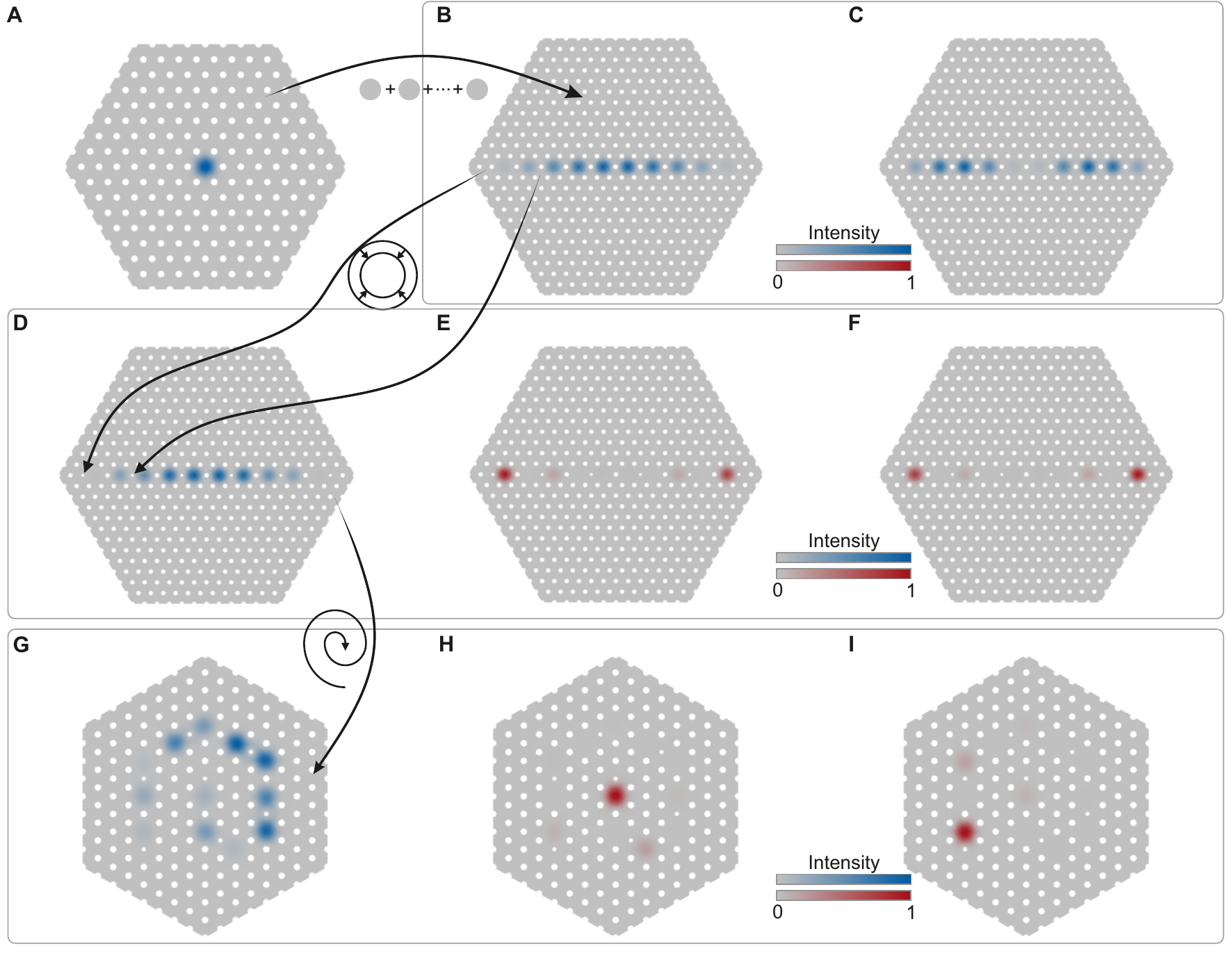}
\caption{\label{fig:design}
\textbf{Explanation of design process}. \textbf{A}, Simulation of an endlessly single mode PCF with a single light-guiding core. Placing multiple cores in the same cladding allows the fibre to support composite supermodes. All of these supermodes have profiles distributed across the entire chain (\textbf{B} and \textbf{C}). Shrinking the air holes between every other pair of cores creates a two-core unit cell with SSH-like coupling. \textbf{D}, Topological bulk mode supported by the multi-core chain with alternating air holes.  \textbf{E}--\textbf{F}, Two degenerate topological edge modes supported by the fibre. The fibre cross-section in \textbf{D}--\textbf{F} can be reduced in size at fixed chain length by following the symmetries of the fibre. \textbf{G}, Topological bulk mode after twisting the chain into a spiral. \textbf{H}--\textbf{I}, Topological edge modes supported by the spiral chain.
}
\end{figure*}

\begin{figure}[tbp]
\centering
\includegraphics[width=\columnwidth]{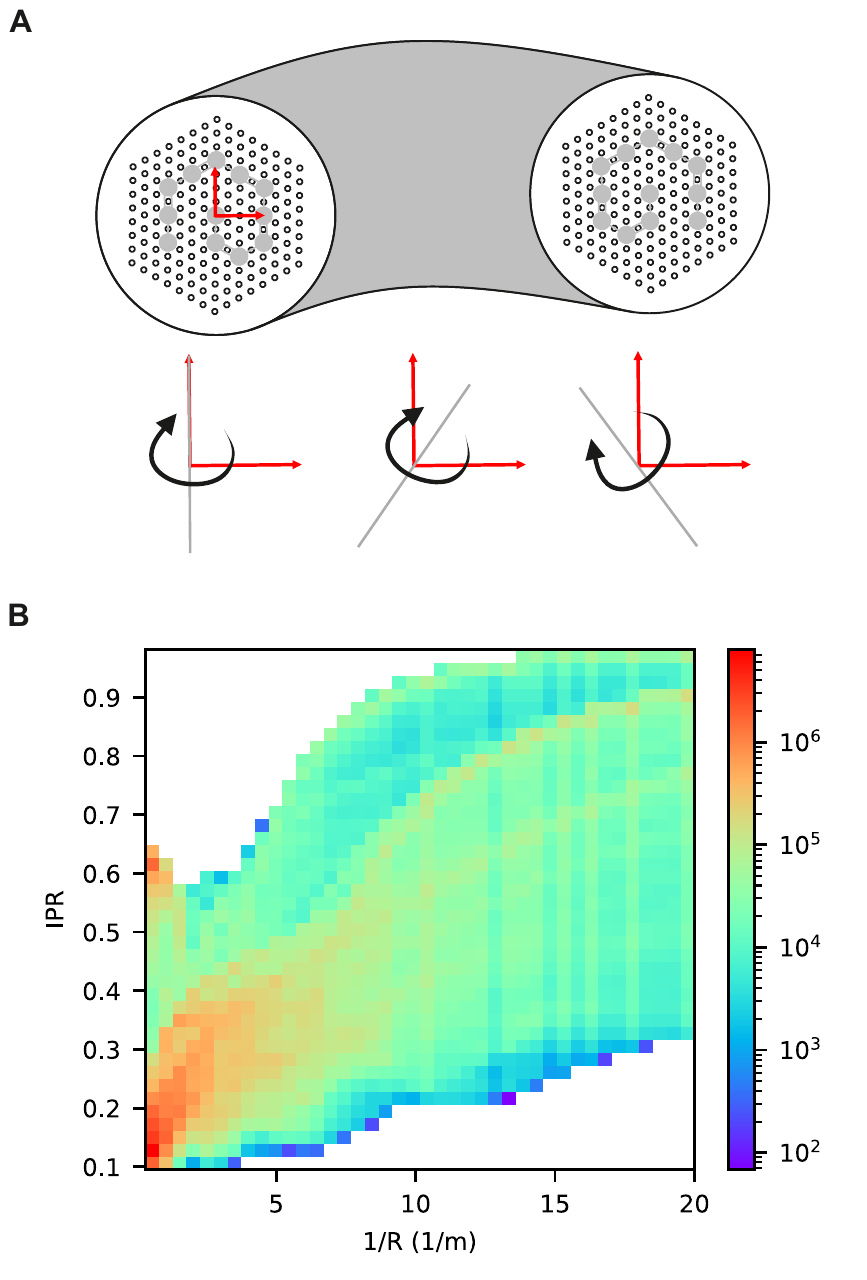}
\caption{\label{fig:theorybend}
\textbf{Effects of bending around a random axis}. To determine the connection between on-site disorder and random bending, we first find the distance of every core from the random bend axis. \textbf{A}, Shows three examples of random bend axis (grey) relative to the fibre cross-section (red) in the lower left, middle, and right images. \textbf{B}, Random bending effects on IPR for 40,000 different bend angles at 200 different bend radii are plotted using a 2D histogram. The upper (lower) red spot towards the left side of the graph corresponds to the edge (bulk) modes. This shows that the modes can be separated into two distinct populations. Only edge modes were selected to be plotted in Fig.~3.}
\end{figure}



\end{document}